\documentclass[rmp,aps,twocolumn]{revtex4}

\newcommand{\qbar}{q}
\newcommand{\bq}{\begin{quotation}\noindent}
\newcommand{\eq}{\end{quotation}}
\newcommand{\be}{\begin{equation}}
\newcommand{\ee}{\end{equation}}
\newcommand{\bea}{\begin{eqnarray}}
\newcommand{\eea}{\end{eqnarray}}
\newcommand{\bc}{\begin{center}}
\newcommand{\ec}{\end{center}}
\def\tr{{\rm tr}\,}
\def\drangle{\rangle\!\rangle}
\def\dlangle{\langle\!\langle}

\newcommand{\boxedeqn}[1]{%
  \[\fbox{%
      \addtolength{\linewidth}{-2\fboxsep}%
      \addtolength{\linewidth}{-2\fboxrule}%
      \begin{minipage}{\linewidth}%
      $$#1$$%
      \end{minipage}%
    }\]%
}

\newcommand{\veec}[3]{\left(%
\begin{array}{r}{\!\!#1\!\!}\\{\!\!#2\!\!}\\{\!\!#3\!\!}\end{array}\right)}

\newtheorem{assump}{Assumption}
\newtheorem{resump}{Resumption}

\usepackage{graphicx}
\usepackage{indentfirst}
\usepackage{amssymb}

\begin{document}

\title{Quantum-Bayesian Coherence: The No-Nonsense Version}

\author{Christopher A. Fuchs$^{\dagger,\ddagger}$ and R\"udiger Schack$^{\sharp,\ddagger}$
\medskip
\\
\small
$^\dagger$Perimeter Institute for Theoretical Physics
\\
\small
%31 Caroline St.\ North,
Waterloo, Ontario N2L 2Y5, Canada
\medskip\\
\small
$^\sharp$Department of Mathematics, Royal Holloway, University of London
\\
\small
Egham, Surrey TW20 0EX, United Kingdom
\medskip\\
\small
$^\ddagger$Stellenbosch Institute for Advanced Study (STIAS)\\
\small
Wallenberg Research Centre at Stellenbosch University\\
\small
%Marais Street,
Stellenbosch 7600, South Africa
}

\date{\today}

\begin{abstract}
In the Quantum-Bayesian interpretation of quantum theory (or QBism), the Born Rule
cannot be interpreted as a rule for setting measurement-outcome probabilities
from an {\it objective\/} quantum state.  But if not, what is the role of the
rule?  In this paper, we argue that it should be seen as an empirical addition
to Bayesian reasoning itself.  Particularly, we show how to view the Born Rule
as a normative rule in addition to usual Dutch-book coherence.  It is a rule
that takes into account how one should assign probabilities to the consequences
of various intended measurements on a physical system, but explicitly in terms
of prior probabilities for and conditional probabilities consequent upon the
imagined outcomes of a special {\it counterfactual\/} reference
measurement. This interpretation is seen particularly clearly by representing
quantum states in terms of probabilities for the outcomes of a fixed, fiducial
symmetric informationally complete (SIC) measurement.  We further explore the
extent to which the general form of the new normative rule implies the full
state-space structure of quantum mechanics.
\end{abstract}

\maketitle

\tableofcontents

\section{Introduction: Unperformed Measurements Have No Outcomes}

\begin{flushright}
\baselineskip=13pt
\parbox{2.8in}{\baselineskip=13pt\small
We choose to examine a phenomenon which is impossible, {\it absolutely\/}
impossible, to explain in any classical way, and which has in it the heart of
quantum mechanics.  In reality, it contains the {\it only\/} mystery.  We
cannot make the mystery go away by ``explaining'' how it works.  We will just
{\it tell\/} you how it works.  In telling you how it works we will have told
you about the basic peculiarities of all quantum mechanics.}
\smallskip\\
\small --- R. P. Feynman, 1964
\end{flushright}

These words come from the opening chapter on quantum mechanics in Richard Feynman's famous {\sl Feynman Lectures on Physics} \cite{Feynman64}.  With them he plunged into a discussion of
the double-slit experiment using {\it individual\/} electrons.  Imagine if you will, however, someone well-versed
in the quantum foundations debates of the last 30 years---since the Aspect
experiment say \cite{Aspect82}---yet naively unaware of when Feynman wrote this.
What might he conclude that Feynman was talking about?  Would it be the
double-slit experiment?  Probably not.  To the modern mindset, a good guess
would be that Feynman was talking about something to do with quantum entanglement or
Bell-inequality violations.  In the history of foundational thinking, the
double-slit experiment has fallen by the wayside.

So, what is it that quantum entanglement teaches us---via
EPR-type considerations and Bell-inequality violations---that
the double-slit experiment does not?  A common answer is
that quantum mechanics does not admit a ``local hidden variable''
formulation.\footnote{Too quick and dirty, some would say
  \cite{Norsen06}. However, the conclusion drawn there---that a Bell inequality
  violation implies the failure of locality, full stop---is based (in part) on
  taking the EPR criterion of reality or variants of it as sacrosanct. As will become clear in
  this paper, we do not take it so.}  By this one usually means the conjunction of two statements
\cite{Bell64,Bell81}: 1) that experiments in one region of spacetime
cannot instantaneously affect matters of fact at far away regions of spacetime,
and 2) that there exist ``hidden variables'' that in some way
``explain'' measured values or their probabilities.
Bell-inequality violations imply that one or the other or
some combination of both these statements fails.  This, many would say, is the
deepest ``mystery'' of quantum mechanics.

This mystery has two sides.  It seems the majority
of physicists who care about these things think it is locality (condition 1
above) that has to be abandoned through the force of the experimentally observed Bell-inequality violations---i.e., they
think there really are ``spooky actions at a distance.''\footnote{Indeed,
  it flavors almost everything they think of quantum mechanics, including the
  {\it interpretation\/} of the toy models they use to
  better understand the theory. Take the recent flurry of work on
  Popescu-Rohrlich boxes \cite{Popescu94}.  These are imaginary devices that
  give rise to greater-than-quantum violations of various Bell inequalities.
  Importantly, another common name for these devices is the term ``nonlocal
  boxes'' \cite{Barrett05}.  Their exact definition comes via the magnitude of
  a Bell-inequality violation---which entails the non-pre-existence of values
  or a violation of locality or both---but the commonly used name opts only to
  recognize nonlocality.  They're not called no-hidden-variable boxes, for instance.
  The nomenclature is psychologically telling.}  Yet, there is a minority
that thinks the abandonment of condition 2 is the more warranted conclusion
\cite{Peres78,Wheeler82,Zeilinger96,Mermin99,Zukowski05,Plotnitsky06,D'Ariano08,Demopoulos08}.
Among these are the {\it Quantum Bayesians\/}
\cite{Fuchs01,Caves02,Schack01,Fuchs02,Fuchs03,Schack04,Fuchs04,Caves07,Appleby05a,Appleby05b,Timpson08,Fuchs10,Fuchs12,Mermin12}.\footnote{For
  alternative developments of several Bayesian-inspired ideas in quantum
  mechanics, see
   \cite{Baez03,Youssef01,Pitowsky03,Pitowsky05,Srednicki05,Caticha06,Leifer06a,Leifer06b,Mana07,Rau07,Goyal08,Warmuth09,Bub11,Leifer12}.
  We leave keep these citations separate from the remark above because of various distinctions within each
  from what we are calling Quantum Bayesianism---these distinctions range from 1) the
  particular strains of Bayesianism each adopts, to 2) whether quantum
  mechanics is a {\it generalized\/} probability theory or rather simply an
  application within Bayesian probability per se, to 3) the level of the
  agent's involvement in bringing about the outcomes of quantum
  measurements. There are nonetheless sometimes striking kinships between the
  ideas of these papers and the effort here, and the papers are well worth
  studying.}  Giving up on hidden variables implies in particular that measured
values do not {\it pre-exist\/} the act of measurement. A measurement does
not merely ``read off'' the values, but enacts or creates them by the
process itself. In a slogan inspired by Asher
Peres \cite{Peres78}, ``unperformed measurements have no outcomes.''

Among the various arguments the Quantum Bayesians use to come to
this conclusion, not least in importance is a thoroughgoing personalist account
of {\it all\/} probabilities
\cite{Ramsey26,DeFinetti31,Savage54,DeFinetti90,Bernardo94,Jeffrey04}---where
the ``all'' in this sentence includes probabilities for quantum measurement
outcomes and even the probability-1 assignments among these \cite{Caves07}.
From the Quantum-Bayesian point of view, this is the only sound interpretation
of probability.  Moreover, this move for quantum probabilities frees up the
quantum state from any objectivist obligations.  In so doing it wipes out the
mystery of quantum-state-change at a distance
\cite{Einstein51,Fuchs00,Timpson08} and much of the mystery of wave function collapse as
well \cite{Fuchs02,Fuchs10,Struggles}.

But what does all this have to do with Feynman?  Apparently Feynman too saw
something of a truth in the idea that ``unperformed measurements have no
outcomes.''  Yet, he did so because of considerations to do with the
double-slit experiment.  Later in the lecture he wrote,
\begin{quotation}
Is it true, or is it {\it not\/} true that the electron either goes through
hole 1 or it goes through hole 2?  The only answer that can be given is that we
have found from experiment that there is a certain special way that we have to
think in order that we do not get into inconsistencies.  What we must say (to
avoid making wrong predictions) is the following.  If one looks at the holes
or, more accurately, if one has a piece of apparatus which is capable of
determining whether the electrons go through hole 1 or hole 2, then one {\it
  can\/} say that it goes either through hole 1 or hole 2.  {\it But}, when one
does {\it not\/} try to tell which way the electron goes, when there is nothing
in the experiment to disturb the electrons, then one may {\it not\/} say that
an electron goes either through hole 1 or hole 2.  If one does say that, and
starts to make any deductions from the statement, he will make errors in the
analysis.  This is the logical tightrope on which we must walk if we wish to
describe nature successfully.
\end{quotation}
Returning to the original quote, we are left with the feeling that this is the
very thing Feynman saw to be the ``basic peculiarity of all quantum
mechanics.''

One should ask though, is his conclusion really compelled by so simple a
phenomenon as the double slit?  How could simple ``interference'' be so
far-reaching in its metaphysical implications?  Water waves interfere and there
is no great mystery there.  Most importantly, the double-slit experiment is a
story of measurement on a single quantum system, whereas the story of EPR and
Bell is that of measurement on {\it two\/} seemingly disconnected systems.

Two systems are introduced for a good reason.  Without the guarantee of
arbitrarily distant parts within the experiment---so that one can conceive of
measurements on one, and draw inferences about the other---what justification
would one have to think that changing the conditions of the experiment (from
one slit closed to both slits open) should {\it not\/} make a deep conceptual
difference to its analysis? Without such a guarantee for underwriting a belief
that some matter of fact stays constant in the consideration of two
experiments, one---it might seem---would be quite justified in responding, ``Of
course, you change an experiment, and you get a different probability
distribution arising from it.  So what?''$\,$\footnote{This is a point Koopman
  \cite{Koopman57} and Ballentine \cite{Ballentine86} seem happy to stop the
  discussion with.  For instance, Ballentine writes, ``One is well advised to
  beware of probability statements of the form, $P(X)$, instead of $P(X|C)$.
  The second argument may be safely omitted only if the conditional event or
  information is clear from the context, and only if it is constant throughout
  the problem.  This is not the case in the double slit experiment. \ldots\ We
  observe from experiment that $P(X|C_3)\ne P(X|C_1)+P(X|C_2)$.  This fact,
  however, has no bearing on the validity of \ldots\ probability
  theory.'' \label{Gobbledygook}} For quite a long time, the authors thought
that Feynman's logical path from {\it example\/} to {\it conclusion}---a conclusion that we
indeed agree with---was simply unwarranted.  The argument just does not seem to
hold to the same stringent standards as Bell-inequality analyses.

However, we have recently started to appreciate that there may be something of
substance in Feynman's argument nonetheless.  It is just not so
easily seen without the proper mindset.  The key point is that the so-called
``interference'' in the example is not in a material field---of course it was
never purported to be---but in something so ethereal as probability itself (a
logical, not a physical, construct).  Most particularly, Feynman makes use of a
beautiful and novel move: He analyzes the probabilities in an experiment that
{\it will be\/} done in terms of the probabilities from experiments that {\it
  won't be\/} done.  He does not simply conditionalize the probabilities to the
two situations and let it go at that.\footnote{See Footnote \ref{Gobbledygook}
  at least once again.}  Rather he tries to see the probabilities in the two
situations as functions of each other. Not functions of a condition, but
functions (or at least relations) of each other.  This is an important point.
There is no necessity that the world give a relation between these probabilities, yet it
does: Quantum mechanics is what makes the link precise.  Feynman seems to have
a grasp on the idea that the essence of the quantum mechanical formalism is to
provide a tool for analyzing the factual in terms of a
counterfactual.\footnote{In his own case, he then develops the formalism
  of amplitudes to mediate between the various probabilities, whereas in this
  paper we will doggedly stick to probabilities, and only probabilities.
  It is only his conceptual point that we want to develop, not his technical apparatus.}

Here is the way Feynman put it in a paper titled, ``The Concept of
Probability in Quantum Mechanics,'' \cite{Feynman51}:
\begin{quotation}
I should say, that in spite of the implication of the title of this talk the
concept of probability is not altered in quantum mechanics.  When I say the
probability of a certain outcome of an experiment is $p$, I mean the
conventional thing, that is, if the experiment is repeated many times one
expects that the fraction of those which give the outcome in question is
roughly $p$.  I will not be at all concerned with analyzing or defining this
concept in more detail, for no departure from the concept used in classical
statistics is required.

What is changed, and changed radically, is the method of calculating
probabilities.
%The effect of this change is greatest when dealing with objects of %atomic dimensions.
\end{quotation}
Far be it from us to completely agree with everything in this quote.  For
instance, the concept of ``probability {\it as\/} long-run frequency'' is
anathema to a Bayesian of any variety \cite{Good83}.\footnote{It is worth
pointing out, however, that Feynman was not always consistently a frequentist
in his thinking about probability. For instance, in the
{\sl Lectures on Physics}, chapter I-6, it says \cite{Feynman64}
\begin{quotation} An experimental
physicist usually says that an ``experimentally
determined'' probability has an ``error,'' and writes
$$P(H) = \frac{N_H}{N} \pm \frac{1}{2\sqrt{N}}\;.$$
There is an implication in such an expression that there {\em is} a
``true'' or ``correct'' probability which {\em could} be computed if we
knew enough, and that the observation may be in ``error'' due to a
fluctuation.  There is, however, no way to make such thinking logically
consistent.  It is probably better to realize that the probability concept
is in a sense subjective, that it is always based on uncertain knowledge,
and that its quantitative evaluation is subject to change as we obtain
more information.\end{quotation}}  And B.~O. Koopman
\cite{Koopman57} is surely right when he says, \bq \indent Ever since the
advent of modern quantum mechanics in the late 1920's, the idea has been
prevalent that the classical laws of probability cease, in some sense, to be
valid in the new theory. More or less explicit statements to this effect have
been made in large number and by many of the most eminent workers in the new
physics.\footnote{In fact, Koopman is speaking directly of Feynman here.
  Moreover, both he and Ballentine \cite{Ballentine86} have criticized Feynman
  on the same point: That with his choice of the word ``changed'' in the last
  quote, Feynman implicates himself in not recognizing that the conditions of
  the three contemplated experiments are distinct and, hence, in not
  conditionalizing his probabilities appropriately.  Thus---Koopman and
  Ballentine say---it is no wonder Feynman thinks he sees a violation of the
  laws of probability. In the authors' opinion however, Koopman and Ballentine
  are hanging too much on the word ``changed''---we rather see it as an
  unfortunate choice of wording on Feynman's part.  That he understood that the
  conditions are different in a deep and inescapable way in the three
  contemplated experiments, we feel, is documented well enough in the quote
  above from his 1964 lecture.} \ldots\

Such a thesis is surprising, to say the least, to anyone holding more or less
conventional views regarding the positions of logic, probability, and
experimental science: many of us have been apt---perhaps too naively---to
assume that experiments can lead to conclusions only when worked up by means of
logic and probability, whose laws seem to be on a different level from those of
physical science.  \eq But there is a kernel of truth here that should not be
dismissed in spite of Feynman's diversion to frequentism and his in-the-end undefensible choice of
the word ``changed.'' Paraphrasing Feynman,
\begin{center}
\parbox{2.6in}{\it The concept of probability is not altered in quantum
  mechanics (it is personalistic Bayes\-ian probability).  What is
  radical is the recipe it gives for calculating new probabilities from old.}
\end{center}

For, quantum mechanics---we plan to show in this paper---gives a resource that
raw Bayesian probability theory does not: It gives a rule for forming
probabilities for the outcomes of {\it factualizable\,}\footnote{We coin this
  term because it stands as a better counterpoint to the term
  ``counterfactual'' than the term ``actualizable'' seems to.  We also want
  to capture the following idea a little more plainly: Both measurements being
  spoken of here are only potential measurements---it is just that one will
  always be considered in the imaginary realm, whereas the other may one day
  become a fact of the matter if it is actually performed.} experiments
(experiments that may actually be performed) from the probabilities one assigns
for the outcomes of a designated {\it counterfactual\/} experiment (an
experiment only imagined, and though possible to do, never actually performed).
So, yes, unperformed measurements have no outcomes as Peres expressed
nicely; nonetheless, imagining their performance can aid in analyzing the
probabilities one {\it ought to\/} assign for an experiment that may
factually be performed.  Putting it more carefully than Feynman: Quantum mechanics
does not provide a radical change to the method of calculating probabilities;
it provides rather an empirical {\it addition\/} to the laws of Bayesian
probability.

In this paper, we offer a modernization and Bayesianification of Feynman's consideration by making intimate use of a potential representation of quantum states\footnote{We say ``potential'' because so far the representation has been seen to exist only for finite dimensional quantum systems with dimension $\le100$.  More will be said on this in Section III.}
unknown in his time:
It is one based on SICs or symmetric informationally complete observables
\cite{Zauner99,Caves99,Renes04,Fuchs04b,Appleby05,ApplebyDangFuchs}.  The goal
is to make it more transparent than ever that the content of the Born Rule
is not that it gives a procedure for {\it setting\/} probabilities (from some
independent entity called ``the quantum state''), but that it represents a
``method of calculating probabilities,'' new ones from old.

That this must be the meaning of the Born Rule more generally for
Quantum Bayesianism has been argued from several
angles by the authors before \cite{Caves07,Struggles}.  For instance, in
\cite{Caves07}, we put it this way:
\bq
\indent We have emphasized
that one of the arguments often repeated to justify that quan\-tum-mechanical
probabilities are objective, rather than subjective, is that they are
``determined by physical law.''  But, what can this mean?  Inevitably, what is
being invoked is an idea that quantum states $|\psi\rangle$ have an existence
independent of the probabilities they give rise to through the Born Rule,
$p(i)=\langle\psi| E_i | \psi\rangle$. From the Bayesian perspective, however,
these expressions are not independent at all, and what we have argued \ldots\ is that quantum states are every bit as subjective as any Bayesian
probability.  What then is the role of the Born Rule?  Can it be dispensed with
completely?

It seems no, it cannot be dispensed with, even from the Bayesian perspective.
But its significance is different than in other developments of quantum
foundations: The Born Rule is not a rule for {\it setting\/} probabilities, but
rather a rule for {\it transforming\/} or {\it relating\/} them.

For instance, take a complete set of $d+1$ observables $O^k$, $k=1,\ldots,d+1$,
for a Hilbert space of dimension $d$.  Subjectively setting probabilities for
the $d$ outcomes of each such measurement uniquely determines a quantum state
$|\psi\rangle$ (via inverting the Born Rule).  Thus, as concerns probabilities
for the outcomes of any other quantum measurements, there can be no more
freedom.  All further probabilities are obtained through linear transformations
of the originals.  In this way, the role of the Born Rule can be seen as having
something of the flavor of Dutch-book coherence, but with an empirical content
added on top of bare, law-of-thought probability theory: An agent interacting
with the quantum world would be wise to adjust his probabilities for the
outcomes of various measurements to those of quantum form if he wants to avoid
devastating consequences. The role of physical law---i.e., the assumption that
the world is made of quantum mechanical stuff---is codified in how measurement
probabilities are related, not how they are set.  \eq What is new in the
present paper is the emphasis on a single designated observable for the
counterfactual thinking, as well as a detailed exploration of the rule for
combining probabilities in this picture.  Particularly, we will see that a
significant part of the structure of quantum-state space arises from the
consistency of that rule---a single formula we designate the {\it urgleichung\/} (German for ``primal equation'').  The urgleichung is the stand-in and correction in our
context for Feynman's not-quite-right but nonetheless suggestive sentence, ``What is
changed, and changed radically, is the method of calculating probabilities.''

Returning to the point we started our discussion with, the one about
interference going by the wayside in quantum foundations, we should say the
following.  To the extent that the full formalism of quantum mechanics can be
re-derived from a simple Feynman-style scenario---even if not the double-slit
experiment per se, but nonetheless one wherein probabilities for the outcomes
of factualizable experiments are obtained from probabilities in a family of designated
counterfactual ones---that scenario must indeed express the essence of quantum
mechanics.  For if these considerations give rise to the full formalism of the
theory (Hilbert spaces, positive operators, the possibility of tensor-product
decompositions, etc.), they must give rise to entanglement, Bell-inequality
violations, and Kochen-Specker `paradoxes' as well: These will be established
as corollaries to the formalism.  Hidden in these
ostensibly different considerations would be every mystery and every `paradox' of quantum mechanics.
And if this is truly the case, who could say that the simple
scenario does not carry in it the essence of quantum mechanics?  From our point
of view, it goes without saying that the exploration of quantum mechanics'
ability to engender Bell-inequality violations and Kochen-Specker theorems is
an immensely instructive activity for sorting out the implications of the
theory.  Nonetheless, one should not lose sight of the potential loss of
understanding that can be incurred by confusing a corollary with a postulate.
In this sense, Feynman may well have had the right intuition.

\subsection{Outline of the Paper}

The plan of the paper is as follows.  In Section II, we review the personalist
Bayesian account of probability, showing how some Dutch-book arguments work,
and emphasizing a point we have not seen emphasized before: Bayes' rule and the
Law of Total Probability, Eqs.\ (\ref{BayesBoy}) and (\ref{TotalProb}), are not
necessities in a Bayesian account of probability.  These rules are enforceable
when there is a {\it conditional lottery\/} in the picture that can be
gambled upon.  But when there is no such lottery, the rules hold no force---without
a conditional lottery there is nothing in Dutch-book coherence itself that can
be used to compel the rules.

In Section III, we review the notion of a SIC (symmetric informationally complete
observable), and show a sense in which it is a very special
measurement. Most importantly we delineate the full structure of quantum-state
space in SIC terms. It turns out that, by making use of a SIC instead of any
other informationally complete measurement, the formalism becomes uniquely
simple and compact. We also show that unitary time evolution, when written in
SIC terms, looks (formally at least) almost identical to classical stochastic evolution.

Section IV contains the heart of the paper.  In it, we introduce the idea of
thinking of an imaginary (counterfactual) SIC behind all quantum measurements,
so as to give an imaginary conditional lottery with which to define conditional
probabilities.  We then show how to write the Born Rule in these terms, and
find it to appear strikingly similar to the Law of Total Probability,
Eq.~(\ref{TotalProb}).  We then note how this move in interpretation is
radically different from the one offered by the followers of ``objective
chance'' in the Lewisian sense \cite{Lewis86a,Lewis86b}.

In Section V, we show that one can derive some of the features of
quantum-state space by taking this modified or {\it Quantum\/} Law of Total
Probability as a postulate.  Particularly, we show that with a
small number of further assumptions, it gives rise to a generalized Bloch sphere and seems to define an
underlying ``dimensionality'' for a system that matches the one given by its
quantum mechanical state space. We also demonstrate other features
of the geometry these considerations give rise to.

In Section VI, we step back further in our considerations and explore the
extent to which the particular constants $d^2$, $d+1$, and $\frac{1}{d}$ in our
Quantum Law of Total Probability, Eq.~(\ref{PrettyBoyBabyThisBeLloyd}), can arise from more elementary
considerations.  This section is a preliminary attempt to understand the origin
of the equation treated simply as a postulate in Section V.

In Section VII we give a brief discussion of where we stand at this stage of
research.  Finally in Section VIII, we close the paper by discussing how our work
is still far from done: Hilbert space, from a Quantum-Bayesian view, has not
yet been derived, only indicated.  Nonetheless the progress made here gives us
hope that we are inching our way toward a formal expression of the ontology
underlying a Quantum-Bayesian vision of quantum mechanics: It has
to do with the Peres slogan ``unperformed measurements have no outcomes,'' but tempered with a kind of `realism' that he
probably would not have accepted forthrightly.\footnote{We say this because of
  Peres's openly acknowledged positivist tendencies.  See Chapters 22 and 23 in
  \cite{Fuchs01} where Peres would sometimes call himself a ``recalcitrant
  positivist.''  Also see the opening remarks of \cite{Peres03} for a good
  laugh.}  On the other hand, it is not a `realism' that we expect to be
immediately accepted by most modern philosophers of science
either.\footnote{See \cite{Nagel89,Dennett04,Price97} for introductions to the ``view from nowhere'' and``view from nowhen'' {\it
    weltanschauungen}.}  This is because it is already clear
that whatever it will ultimately turn out to be, it is based on a) a rejection
of the ontology of the block universe \cite{James1882,James1884,James1910}, and
b) a rejection of the ontology of the detached observer
\cite{Pauli94,Laurikainen88,Gieser05}.  The `realism' in vogue in philosophy-of-science circles, which makes heavy use of both these elements, is, as Wolfgang Pauli once said,  ``too narrow a concept'' for our purposes \cite{Pauli94}.  Reality, the stuff of which the world is made, the stuff that was here before agents and observers, is more interesting than that.

\section{Personalist Bayesian Probability}

From the Bayesian point of view, probability is degree of belief as measured by
action.  More precisely, we say one has (explicitly or implicitly) assigned a
probability $p(A)$ to an event $A$ if, before knowing the value of $A$, one is
willing to either {\it buy\/} or {\it sell\/} a lottery ticket of the form
\begin{center}
\parbox{2.0in}{\boxedeqn{\raisebox{1.5\height}{\mbox{Worth \$1 if $A$.}}}}
\end{center}
for an amount $\$ p(A)$.\footnote{In other words, the personalist Bayesian
  agent regards $\$p(A)$ as the {\it fair\/} price of the lottery ticket. He
  would regard it as advantageous to buy it for any price less than $\$p(A)$,
  or to sell it for any price greater than $\$p(A)$.}
The personalist Bayesian position adds only that this
is the full meaning of probability; it is nothing more and nothing less than
this definition.  Particularly, nothing intrinsic to the event or proposition
$A$ can help declare $p(A)$ right or wrong, or more or less rational.  The
value $p(A)$ is solely a statement about the agent who assigns it.

Nonetheless, even for a personalist Bayesian, probabilities do not wave in the
wind.  Probabilities are held together by a normative principle: That whenever
an agent declares probabilities for various events---say $A$, $\neg B$ (``not
$B$''), $A\vee B$ (``$A$ or $B$''), $A\wedge C$ (``$A$ and $C$''), etc.---he
should {\it strive\/} to never gamble (i.e., buy and sell lottery tickets) so
as to incur what he believes will be a sure loss.  This normative principle is
known as Dutch-book coherence.  And from it, one can derive the usual calculus
of probability theory.

This package of views about probability (that in value it is personal, but that
in function it is akin to the laws of logic) had its origin in the mid-1920s
and early 1930s with the work of F.~P. Ramsey \cite{Ramsey26} and B. de Finetti
\cite{DeFinetti31}.  J. M. Keynes characterizes Ramsey's position
succinctly \cite{Keynes51}:
\begin{quotation}
\noindent [Ramsey] succeeds in showing that the calculus of
probabilities simply amounts to a set of rules for ensuring that the
system of degrees of belief which we hold shall be a {\it
consistent\/} system.  Thus the calculus of probabilities belongs to
formal logic.  But the basis of our degrees of belief---or the {\it a
priori}, as they used to be called---is part of our human outfit,
perhaps given us merely by natural selection, analogous to our
perceptions and our memories rather than to formal logic.
\end{quotation}
And B.~O. Koopman writes
\cite{Koopman40}: \bq \indent The intuitive thesis in probability holds that
both in its meanings and in the laws which it obeys, probability derives
directly from the intuition, and is prior to objective experience; it holds
that it is for experience to be interpreted in terms of probability and not for
probability to be interpreted in terms of experience \ldots
\eq

Let us go through some of the derivation of the probability calculus from
Dutch-book coherence so that we may better make a point concerning quantum
mechanics afterward.\footnote{Here we basically follow the development in
  Richard Jeffrey's posthumously published book {\sl Subjective Probability,
    The Real Thing} \cite{Jeffrey04,Skyrms87}, but with our own emphasis.}
First we establish that our normative principle requires $0\le P(A)\le 1$.  For
suppose $P(A)<0$.  This means an agent will sell a ticket for negative
money---i.e., he will pay someone $\$ p(A)$ to take the ticket off his hands.
Regardless of whether $A$ occurs or not, the agent will then be sure he will
lose money.  This violates the normative principle.  Now, take the case
$P(A)>1$.  This means the agent will buy a ticket for more than it is worth
even in the best case---again a sure loss for him and a violation of the
normative principle.  So, probability in the sense of ticket pricing should
obey the usual range of values.

Now let us establish the probability sum rule.  Suppose our agent believes two
events $A$ and $B$ to be mutually exclusive---i.e., he is sure that if $A$
occurs, $B$ will not, or if $B$ occurs, $A$ will not. We can contemplate three
distinct lottery tickets:
\begin{center}
\parbox{2.0in}{\boxedeqn{\raisebox{1.5\height}{\mbox{Worth \$1 if $A\vee B$.}}}}
\end{center}
and
\begin{center}
\parbox{1.2in}{\boxedeqn{\raisebox{1.5\height}{\mbox{Worth \$1 if $A$.}}}}
%\end{center}
\qquad\quad
%\begin{center}
\parbox{1.2in}{\boxedeqn{\raisebox{1.5\height}{\mbox{Worth \$1 if $B$.}}}}
\end{center}
Clearly the value of the first ticket should be the same as the total value of the
other two.  For instance, suppose an agent had set $P(A\vee B)$, $P(A)$ and
$P(B)$ such that $P(A\vee B) > P(A) + P(B)$.  Then---by definition---when
confronted with a seller of the first ticket, he must be willing to buy it, and
when confronted with a buyer of the other two tickets, he must be willing to
sell them.  But then the agent's initial balance sheet would be negative: $-\$
P(A\vee B) + \$P(A) + \$P(B)<\$0$.  And whether $A$ or $B$ or neither event
occurs, it would not improve his finances: If a dollar flows in (because of the
bought ticket), it will also flow out (because of the agent's responsibilities
for the sold tickets), and still the balance sheet is negative.  The agent is
sure of a loss.  A similar argument goes through if the agent had set his
ticket prices so that $P(A\vee B) < P(A) + P(B)$.  Thus whatever values are
set, the normative principle prescribes that it had better be the case that
$P(A\vee B) = P(A) + P(B)$.

Consider now the following lottery ticket of a slightly different structure:
\begin{center}
\parbox{2.0in}{\boxedeqn{\raisebox{1.5\height}{\mbox{Worth \,{\$}$\frac{m}{n}$\, if $A$.}}}}
\end{center}
where $m\le n$ are integers. Does Dutch-book coherence say anything about the
value of this ticket in comparison to the value of the standard ticket---i.e.,
one worth \$1 if $A$?  It does.  An argument quite like the one above dictates
that it should be valued {\$}$\frac{m}{n}P(A)$.  If a real number $\alpha$ were
in place of the $\frac{m}{n}$ a similar result follows from a limiting
argument.

Now we come to the most interesting and important case.  Bayesian probability
is not called by its name for lack of a good reason.  A key rule in probability
theory is Bayes' rule relating joint to conditional probabilities: \be
p(A\wedge B)=p(A)p(B|A)\;.
\label{BayesBoy}
\ee Like the rest of the structure of probability theory within the Bayesian
conception, this rule must arise from an application of Dutch-book coherence.
What is that application?

The only way anyone has seen how to do it is to introduce the idea of a {\it
  conditional lottery} \cite{Kyburg80}.  In such a lottery, the value of the
event $A$ is revealed to the agent first.  If $A$ obtains, the lottery proceeds
to the revelation of event $B$, and finally all monies are settled up.  If on
the other hand $\neg A$ obtains, the remainder of the lottery is called off,
and the monies put down for any ``conditional tickets'' are returned.  That is
to say, the meaning of $p(B|A)$ is taken to be the price {\$}$p(B|A)$ at which
one is willing to {\it buy\/} or {\it sell\/} a lottery ticket of the following
form:

\vspace{-0.15in}
\begin{center}
\parbox{3.2in}{\boxedeqn{\raisebox{1.5\height}{\mbox{Worth \,{\$}1 if $A\wedge B$.
\quad But return price if $\neg A$.
}}}}
\end{center}
for price {\$}$p(B|A)\;$.
Explicitly inserting the definition of $p(B|A)$, this becomes:

\vspace{-0.15in}
\begin{center}
\parbox{3.3in}{\boxedeqn{\raisebox{1.5\height}{\mbox{Worth \,{\$}1 if $A\wedge B$.
\quad But return {\$}$p(B|A)$ if $\neg A$.
}}}}
\end{center}
Now comes the coherence argument.  For, if you think about it, the price for
this ticket had better be the same as the total price for these two tickets:
\begin{center}
\parbox{2.0in}{\boxedeqn{\raisebox{1.5\height}{\mbox{Worth \$1 if $A\wedge B$.}}}}
\qquad\quad
\parbox{2.0in}{\boxedeqn{\raisebox{1.5\height}{\mbox{Worth {\$}$p(B|A)$ if $\neg A$.}}}}
\end{center}
That is to say, to guard against a sure loss, we must have \bea p(B|A) &=&
p(A\wedge B) + p(B|A)p(\neg A) \nonumber \\ &=& p(A\wedge B) + p(B|A) - p(B|A)
p(A)\;.  \eea Consequently, Eq.~(\ref{BayesBoy}) should hold whenever there is
a conditional lottery under consideration.

\subsection{When a Conditional Lottery Is Not Without Consequence}

\label{OhBoy}

But what if the conditional lottery is called off because the draw that was to
give rise to the event $A$ does not take place? In this case the probabilities
$p(A)$ and $p(B|A)$ refer to a counterfactual, and there is no reason to assume
the validity of Eq.~(\ref{BayesBoy}).

It is worth investigating the idea of counterfactuals in some more detail.
Suppose an agent makes a measurement of a variable $X$ that takes on values
$x$, followed by a measurement of a variable $Y$ with mutually exclusive values
$y$.  A Dutch bookie asks him to commit on various unconditional and
conditional lottery tickets.  What can we say of the probabilities he ought to
ascribe?  A minor variation of the Dutch-book arguments above tells us that
whatever values of $p(x)$, $p(y)$, and $p(y|x)$ he commits to, they ought---if he
is coherent---satisfy the Law of Total Probability:
\be p(y)=\sum_x p(x) p(y|x)\;.
\label{TotalProb}
\ee Imagine now that the $X$ measurement is called off, so there will only be
the $Y$ measurement. Is the agent still normatively committed to buying and
selling $Y$-lottery tickets for the price {\$}$p(y)$ in Eq.~(\ref{TotalProb})
that he originally expressed?  Not at all!  That would clearly be silly in some
situations, and no clear-headed Bayesian would ever imagine otherwise.  The
action bringing about the result of the $X$ measurement might so change the
situation for bringing about $Y$ that he simply would not gamble on it in the
same way.  To hold fast to the {\$}$p(y)$ valuation of a $Y$-lottery ticket,
then, is not a necessity enforced by coherence, but a judgment that might or
might not be the right thing to do.

In fact, one might regard holding fast to the initial value {\$}$p(y)$ in
spite of the nullification of the conditional lottery as the formal definition
of precisely what it means to judge an unperformed measurement {\it to have\/}
an outcome.  It means one judges that looking at the value of $X$ is incidental
to the whole affair, and this is reflected in the way one gambles on $Y$
\cite{FuchsSchack2012}.  So, if $q(y)$ represents the probabilities with which the
agent gambles supposing the $X$-lottery nullified, then a formal statement of
the Peresian slogan that the unperformed $X$ measurement had no outcome (i.e.,
measuring $X$ matters, and it matters even if one makes no note of the outcome)
is that \be q(y)\ne p(y)\;.  \ee

Still, one might imagine situations in which even if an agent judges that
equality does not hold for them, he nonetheless judges that $q(y)$ and $p(y)$
should bear a precise relation to each other.  In Section IV, we will show that, in fact, the positive content of the Born Rule as an addition to Bayesianism is to connect the probabilities for two measurements, one factual and one counterfactual, for which Dutch-book coherence alone does not provide a precise relationship.

\section{Expressing Quantum-State Space in Terms of SICs}

Let ${\mathcal H}_d$ be a finite $d$-dimensional Hilbert space associated with some physical system.  A quantum state for the system is usually expressed as a unit-trace positive semi-definite linear operator $\rho\in{\mathcal L}({\mathcal H}_d)\,$. \ However, through the use of a {\it symmetric informationally complete observable\/} (or SIC) as a reference observable, we can find a rather elegant representation of quantum states directly in terms of an associated set of probability distributions.

A SIC is an example of a generalized measurement or positive operator-valued measure (POVM) \cite{PeresBook}.  A POVM is a collection $\{E_i\}$, $i=1,\ldots,n$,
of positive semi-definite operators $E_i$ on ${\mathcal H}_d$ such that \be\sum_i E_i=I\;,\ee where
$n$ is in general unrelated to $d$ and may be larger or smaller than $d$.  Supposing a quantum state $\rho$, the probability of the measurement outcome labeled $i$
is then given by \be p(i)=\tr\rho E_i\;. \label{traceRhoE} \ee The POVMs represent the most
general kinds of quantum measurement that can be made on a system. A von Neumann measurement is a special POVM where
the $E_i$ are mutually orthogonal projection operators. Mathematically, any
POVM can be written as a unitary interaction with an ancillary quantum system,
followed by a von Neumann measurement on the ancillary system \cite{NielsenBook}.

We can provide an injective mapping between the convex set of density operators
and the set of probability distributions\footnote{Please note our pseudo-Dirac
  notation $\|v\drangle$ for vectors in a {\it real\/} vector space of $d^2$
  dimensions.  The relevant probability simplex for us---the one we are mapping
  quantum states $\rho$ to, denoted $\Delta_{d^2}$---is a convex body within
  this linear vector space.  Thus, its points may be expressed with the
  notation $\|p\drangle$ as well.  The choice of a pseudo-Dirac notation for
  probability distributions also emphasizes that one should think of the valid
  $\|p\drangle$ as a direct expression of the set of quantum states.}  \be \|
p\drangle=\Big(p(1), p(2), \ldots, p(d^2)\Big)^{\!\rm T} \ee over $d^2$
outcomes---the probability simplex $\Delta_{d^2}$---by first fixing any so-called
minimal informationally complete fiducial measurement $\{E_i\}$,
$i=1,\ldots,d^2$.  This is a POVM with all the $E_i$ linearly independent.
With respect to such a measurement, the probabilities $p(i)$ for its outcomes
completely specify $\rho$. This follows because the $E_i$ form a basis for
${\mathcal L}({\mathcal H}_d)$, and the probabilities $p(i)=\tr \rho E_i$ can
be viewed as instances of the Hilbert-Schmidt inner product \be (A,B)=\tr
A^\dagger B\;.\ee The quantities $p(i)$ thus merely express the projections of
the vector $\rho$ onto the basis vectors $E_i$.  These projections completely
fix the vector $\rho$.

\begin{figure} %\leavevmode
\begin{center}
\includegraphics[height=2in]{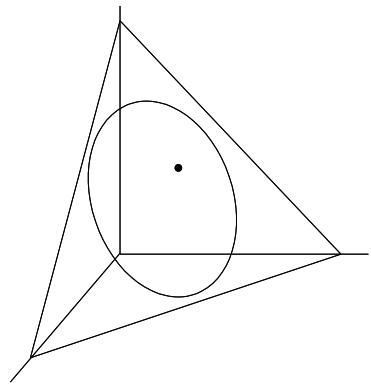}
\bigskip\caption{The planar surface represents the convex set of all
  probability distributions over $d^2$ outcomes---the probability simplex
  $\Delta_{d^2}$.  The probability distributions valid for representing the set
  of quantum states, however, is a smaller convex set within the simplex---here
  depicted as an ellipsoid.  In actual fact, however, the
  convex shape is quite complex.  The choice of a SIC for defining the mapping makes the shape as simple as it can be with respect to the natural
  geometry of the simplex.}
\end{center}
\end{figure}

One can see how to calculate $\rho$ in terms of the vector $\|p\drangle$ in the
following way.  Since the $E_i$ form a basis, there must be some expansion
\be\rho=\sum \alpha_j E_j \;,\ee where the $\alpha_j$ are real numbers making up a
vector $\|\alpha\drangle$. \ Thus, \be p(i)=\sum_j \alpha_j\, \tr E_i E_j\;.
\ee If we let a matrix $M$ be defined by entries \be M_{ij}=\tr E_i E_j\;,\ee this
just becomes \be \|p\drangle = M \|\alpha\drangle\;.  \ee Using the fact that
$M$ is invertible because the $E_i$ are linearly independent, we have finally
\be \|\alpha\drangle = M^{-1} \|p\drangle\;.
\label{YakBoy}
\ee

The most important point of this exercise is that with such a mapping
established, one has every right to think of a quantum state as a probability
distribution {\it full stop}.  In \cite{Fuchs02} it is argued that,
conceptually, it is indeed nothing more.  However, it is important to
note that the mapping $\rho\mapsto \|p\drangle$, though injective, cannot be
surjective.  In other words, only some probability distributions in the simplex
are valid for representing quantum states. A significant part of understanding
quantum mechanics is understanding the range of shapes available under these
mappings \cite{Bengtsson06}.

Particularly, it is important to recognize that informationally complete
measurements abound---they come in all forms and sizes. Hence there is no
unique representation of this variety for quantum states.  A reasonable question
thus follows: What is the best measurement one can use for a mapping
$\rho\mapsto \|p\drangle\,$? \ One would not want to unduly burden the
representation with extra terms and calculations if one does not have to.  For
instance, it would be beautiful if one could take the informationally complete
measurement $\{E_i\}$ so that $M$ is simply a diagonal matrix or even the
identity matrix itself.  Such an extreme simplification, however, is not in the
cards---it cannot be done.

Its failure does, however, point to an interesting direction worthy of development:
If one cannot make $M$ diagonal, one might still want to make $M$ as close to
the identity as possible.  A convenient measure for how far $M$ is from the
identity is the squared Frobenius distance: \bea F &=& \sum_{ij}
\Big(\delta_{ij}-M_{ij}\Big)^2\nonumber \\ &=& \sum_i \Big(1-\tr E_i^2\Big)^2 +
\sum_{i\ne j} \Big(\tr E_i E_j\Big)^2\;.  \eea We can place a lower bound on
this quantity with the help of a special instance of the Schwarz inequality: If
$\lambda_r$ is any set of $n$ nonnegative numbers, then \be \sum_r \lambda_r^2
\ge \frac{1}{n}\!\left(\sum_r \lambda_r\right)^{\! 2}\;, \ee with equality if
and only if $\lambda_1 = \dots = \lambda_n$.  Thus, \bea F &\ge &
\frac{1}{d^2}\left(\sum_i \Big(1-\tr E_i^2\Big)\right)^2 \nonumber\\ && + \;
\frac{1}{d^4-d^2}\left(\sum_{i\ne j} \tr E_i E_j\right)^2\;.  \eea Equality
holds in this if and only if there are constants $m$ and $n$ such that $\tr
E_i^2=m$ for all $i$ and $\tr E_i E_j=n$ for all $i\ne j$.  Since \be d=\tr
I^2=\sum_{ij}\tr E_i E_j = \sum_i \tr E_i^2 + \sum_{i\ne j} \tr E_i E_j\;, \ee
$m$ and $n$ must be related by \be m+(d^2-1)n=\frac{1}{d}\;.  \ee On the other
hand, with these conditions fulfilled the $E_i$ must all have the same trace.
For, \be \tr E_k = \sum_i \tr E_k E_i = m + (d^2-1)n\;.  \ee Consequently \be
\tr E_k = \frac{1}{d}\;. \label{traceEk} \ee Now, how large can $m$ be?  Take a positive
semi-definite matrix $A$ with $\tr A=1$ and eigenvalues $\lambda_i$.  Then
$\lambda_i\le 1$, and clearly $\tr A^2\le \tr A$ with equality if and only if
the largest $\lambda_i$ is equal to 1.  Hence, $d E_k$ will give the largest
allowed value $m$ if $E_i = \frac{1}{d} \Pi_i$, where \be\Pi_i
=|\psi_i\rangle\langle \psi_i|\;, \label{psipsi} \ee for some rank-1 projection operator
$\Pi_i$.  If this obtains, $n$ takes the form \be n=\frac{1}{d^2(d+1)}\;.  \ee

In total we have shown that a measurement $\{E_i\}$, $i=1,\ldots,d^2$, will
achieve the best lower bound for $F$ if and only if \be E_i=\frac{1}{d}
\Pi_i \label{Mustard0}\ee with \be \tr \,\Pi_i\Pi_j=\frac{d\delta_{ij}+1}{d+1}\;.
\label{Mustard}
\ee

Significantly, it turns out that measurements of this variety also have the
property of being necessarily informationally complete \cite{Caves99}.
Let us show this for completeness: It is just a matter of proving that the
$E_i$ are linearly independent.  Suppose there are some numbers $\alpha_i$ such
that \be \sum_i \alpha_i E_i = 0\;.
\label{PenScript}
\ee Taking the trace of this equation, we infer that $\sum_i \alpha_i = 0$. Now
multiply Eq.~(\ref{PenScript}) by an arbitrary $E_k$ and take the trace of the
result.  We get, \be \frac{1}{d^2}\sum_i \alpha_i
\frac{d\delta_{ik}+1}{d+1}=0\;.  \ee In other words \be \sum_i \alpha_i
\delta_{ik} = 0\;, \ee which of course implies $\alpha_k=0$.  So the $E_i$ are
linearly independent.

These kinds of measurements are presently a hot topic of study in quantum
information theory, and have come to be known as ``symmetric informationally
complete'' quantum measurements \cite{Caves99}.  As such, the measurement
$\{E_i\}$, the associated set of projection operators $\{\Pi_i\}$, and even the
set of $\{|\psi_i\rangle\}$ are often simply called SIC (pronounced
``seek'').\footnote{This choice of pronunciation is meant to be in accord with
  the ``pedant's pronunciation'' of the Latin adverb {\it sic} \cite{Bennett08}.
  Moreover, it alleviates any potential confusion between the pluralized form
  SICs and the number six in conversation.}  We will adopt that terminology
here.

Here is an example of a SIC in dimension-2, expressed in terms of the Pauli
operators:
\bea
\Pi_1 &=& \frac{1}{2}\left(I+\frac{1}{\sqrt{3}}(\sigma_x+\sigma_y+\sigma_z)\right) \;,\nonumber\\
\Pi_2 &=& \frac{1}{2}\left(I+\frac{1}{\sqrt{3}}(\sigma_x-\sigma_y-\sigma_z)\right) \;,\nonumber\\
\Pi_3 &=& \frac{1}{2}\left(I+\frac{1}{\sqrt{3}}(-\sigma_x-\sigma_y+\sigma_z)\right) \;,\nonumber\\
\Pi_4 &=& \frac{1}{2}\left(I+\frac{1}{\sqrt{3}}(-\sigma_x+\sigma_y-\sigma_z)\right) \;.
\eea
And here is an example of a SIC in dimension-3 \cite{Tabia12}.  Taking $\omega=e^{2\pi i/3}$ to be
a third-root of unity and $\overline{\omega}$ to be its complex conjugate, let
\bea
|\psi_1\rangle={\veec01{-1}}, & |\psi_2\rangle={\veec{-1}01}, & |\psi_3\rangle={\veec1{-1}0} \nonumber\\
|\psi_4\rangle={\veec0\omega{-\overline{\omega}}}, & |\psi_5\rangle={\veec{-1}0{\overline{\omega}}}, & |\psi_6\rangle={\veec1{-\omega}0} \rule{0mm}{10mm} \nonumber\\ |\psi_7\rangle={\veec0{\overline{\omega}}{-\omega}}, & |\psi_8\rangle={\veec{-1}0\omega}, & |\psi_9\rangle={\veec1{-\overline{\omega}}0} \rule{0mm}{10mm} \nonumber\\
\label{Moutard}
\eea
be defined up to normalization.  One can check by quick inspection that (after normalization) these vectors do indeed satisfy Eq.~(\ref{Mustard}).

Do SICs exist for every finite dimension $d$?  Despite many efforts in the last 14 years---see \cite{Caves99,Zauner99,Renes04,Fuchs04b,Appleby05,ApplebyDangFuchs} and particularly the extensive reference lists in \cite{Scott09} and \cite{Appleby12}---no one presently knows.  However, there is a
strong feeling in the community that they do, as analytical proofs have been
obtained for all dimensions $d=2$--$16$, 19, 24, 28, 31, 35, 37, 43, and 48,\footnote{Dimensions 2--5 were published in \cite{Zauner99}. Dimensions 6--16, 24, 28, 35, and 48 are due to M. Grassl in various publications; see \cite{Scott09}. Dimensions 7, 19, 31, 37, and 43 are due to D. M. Appleby, with the latter three as yet unpublished \cite{Appleby12}.} and within a
numerical precision of $10^{-38}$, they have been observed by computational
means \cite{Scott09} in all dimensions $d=2$--$67$.  To lesser numerical precision E. Schnetter has also found SICs in $d=68$--$73$, 75--81, 83, 84, 89, 93, 100 \cite{Schnetter12}.

The SIC structure is ideal
for revealing the essence of the Born Rule as an addition to Dutch-book
coherence. From here out, we will proceed as if SICs do indeed always
exist. The foundational significance of the following technical development
rests on this mathematical conjecture.

Let us spell out in some detail what the set of quantum states written as SIC
probability vectors $\|p\drangle$ looks like. Perhaps the most remarkable thing
about a SIC is the level of simplicity it lends to Eq.~(\ref{YakBoy}).  On top
of the theoretical justification that SICs are as near as possible to an
orthonormal basis, from Eqs.~(\ref{traceRhoE}), (\ref{Mustard0})
and~(\ref{Mustard}) one gets the simple expression \cite{Fuchs04b,Caves02b} \be
\rho=\sum_i \left((d+1)p(i)-\frac{1}{d}\right)\Pi_i\;.
\label{FierceUrgencyOfNow}
\ee In other words, effectively all explicit reference to the matrix $M^{-1}$
disappears from the expression. The components $(d+1)p(i)-\frac{1}{d}$ are
obtained by a universal scalar readjustment from the probabilities $p(i)$.
This will have important implications.

Still, one cannot place just any probability distribution $p(i)$ into Eq.~(\ref{FierceUrgencyOfNow}) and
expect to obtain a positive semi-definite $\rho$.  Only some probability
vectors $\|p\drangle$ are {\it valid\/} ones.  Which ones are they?  For
instance, $p(i)\le \frac{1}{d}$ must be the case, as dictated by
Eqs.~(\ref{traceRhoE}) and~(\ref{traceEk}), and this already restricts the class of
  valid probability assignments.  But there are more requirements than that.

In preparation for characterizing the set of valid probability vectors
$\|p\drangle$, let us note that since the $\Pi_k$ form a basis on the space of
operators, we can define operator multiplication in terms of
them.  This is done by introducing the so-called structure coefficients
$\alpha_{ijk}$ for the algebra: \be \Pi_i\Pi_j=\sum_k \alpha_{ijk}\Pi_k\;.
\label{Hermeneutic}
\ee A couple of properties follow immediately.  Taking the trace of both sides
of Eq.~(\ref{Hermeneutic}), one has \be \sum_k \alpha_{ijk} =
\frac{d\,\delta_{ij}+1}{d+1}\;.
\label{Fritsa}
\ee
Using this, one gets straightforwardly that
\be
\tr\!\Big( \Pi_i\Pi_j\Pi_k\Big)=\frac{1}{d+1}\left(d\,\alpha_{ijk}+
\frac{d\,\delta_{ij}+1}{d+1}\right).
\ee
In other words,
\be
\alpha_{ijk}=\frac{1}{d}\left((d+1)\tr\!\Big(\Pi_i\Pi_j\Pi_k\Big)-
\frac{d\,\delta_{ij}+1}{d+1}\right).
\ee
For the analogue of Eq.~(\ref{Fritsa}) but with summation over the first or second index, one gets,
\be
\sum_i \alpha_{ijk}=d\delta_{jk} \qquad\mbox{and}\qquad\sum_j \alpha_{ijk}=d\delta_{ik}\;.
\ee

With these expressions in hand, one sees a very direct connection between the
structure of the algebra of quantum states when written in operator language
and the structure of quantum states when written in probability-vector
language.  For, the complete convex set of quantum states is fixed by the set
of its extreme points, i.e., the pure quantum states---rank-1 projection
operators.  To characterize this set algebraically, one method is to note that
these are the only hermitian operators satisfying $\rho^2=\rho$.  Using
Eq.~(\ref{FierceUrgencyOfNow}), we find that a quantum state $\|p\drangle$ is
pure if and only if its components satisfy these $d^2$ simultaneous quadratic
equations: \be p(k)=\frac{1}{3}(d+1)\sum_{ij}
\alpha_{ijk}\,p(i)p(j)+\frac{2}{3d(d+1)}\;.
\label{NewFavoredBoy}
\ee Another way to characterize this algebraic variety---an
algebraic variety is defined as the set of solutions of a system of polynomial
equations---is to make use of a theorem of Flammia, Jones, and
Linden \cite{Flammia04,Jones05}: A hermitian operator $A$ is a rank-one
projection operator if and only if $\tr A^2=\tr A^3=1$.\footnote{The theorem is
  nearly trivial to prove once one's attention is drawn to it: Since $A$ is
  hermitian, it has a real eigenvalue spectrum $\lambda_i$. \ From the first
  condition, one has that $\sum_i \lambda_i^2=1$; from the second, $\sum_i
  \lambda_i^3=1$. \ The first condition, however, implies that $|\lambda_i| \le
  1$ for all $i$. \ Consequently $1-\lambda_i\ge0$ for all $i$.  Now taking the
  difference of the two conditions, one sees that $\sum_i
  \lambda_i^2(1-\lambda_i)=0$. \ In order for this to obtain, it must be the
  case that $\lambda_i$ is always $0$ or $1$ exclusively. That there is only
  one nonzero eigenvalue then follows from using the first condition again.
  Thus the theorem is proved.  However, it seems not to have been widely recognized previous to 2004--2005.} \ So
in fact our $d^2$ simultaneous quadratic equations reduce to just two equations
instead, one a quadratic and one a cubic: \be \sum_i p(i)^2=\frac{2}{d(d+1)}
\label{PurePurity}
\ee
and
\be
\sum_{ijk} \alpha_{ijk}\,p(i)p(j)p(k)=\frac{4}{d(d+1)^2}\;.
\label{MonkeyFun}
\ee
Note that Eqs.~(\ref{NewFavoredBoy}) and (\ref{MonkeyFun}) are complex equations, but one could symmetrize them and make them purely real if one wanted to.

There are also some advantages to working out these equations more explicitly
in terms of the completely symmetric 3-index tensor \be
c_{ijk}=\mbox{Re}\;\tr\!\Big(\Pi_i\Pi_j\Pi_k\Big)\;.
\label{TripleSec}
\ee In terms of these quantities, the analogues of Eqs.~(\ref{NewFavoredBoy})
and (\ref{MonkeyFun}) become \be p(k)=\frac{(d+1)^2}{3d}\sum_{ij}
c_{ijk}\,p(i)p(j)-\frac{1}{3d}
\label{MegaMorph}
\ee
and
\be
\sum_{ijk} c_{ijk}\,p(i)p(j)p(k)=\frac{d+7}{(d+1)^3}\;,
\label{ChocolateMouse}
\ee respectively.  The reason for noting this comes from the
simplicity of the $d^2$ matrices $C_k$ with matrix entries $(C_k)_{ij}=c_{ijk}$
from Eq.~(\ref{TripleSec}), which is explored in great detail in \cite{Appleby09}.
To give a flavor of the results, we note
for instance that, for each value of $k$, $C_k$ turns
out to have the form \cite{Appleby09}
\be
C_k=\|m_k\drangle\dlangle
m_k\|+\frac{d}{2(d+1)} Q_k\;,
\label{Magma}
\ee
where the $k$-th vector $\|m_k\drangle$ is defined by \be \|m_k\drangle =
\left( \frac{1}{d+1}, \ldots, 1, \ldots, \frac{1}{d+1}\right)^{\!\rm T}\;, \ee
and $Q_k$ is a $(2d-2)$-dimensional projection operator on the real vector
space embedding the probability simplex $\Delta_{d^2}$.  Furthermore, using
this, one obtains a useful expression for the pure states; they are
probabilities satisfying a simple class of quadratic equations \be
p(k)=d\,p(k)^2 + \frac{1}{2}(d+1)\dlangle p\|Q_k\|p\drangle\;.  \ee

With Eqs.~(\ref{PurePurity}), (\ref{MegaMorph}), and (\ref {ChocolateMouse}) we
have now discussed the extreme points of the convex set of quantum states---the
pure states.  The remainder of the set of quantum states is then constructed by
taking convex combinations of the pure states. This is an implicit expression
of quantum-state space.  But SICs can also help give an explicit
parameterization of the convex set.

We can see this by starting not with density operators, but with ``square
roots'' of density operators.  This is useful because a matrix $\rho$ is
positive semi-definite if and only if it can be written as $\rho=B^2$ for some
hermitian $B$.  Thus, let \be B=\sum_i b_i\Pi_i \ee with $b_i$ a set of real
numbers.  Then, \be \rho=\sum_k\left(\sum_{ij}b_i b_j\alpha_{ijk}\right)\Pi_k
\ee will represent a density operator so long as $\tr \rho=1$. This condition
requires simply that \be \left(\sum_i b_i\right)^2+d\sum_i b_i^2\;=\;d+1\;,
\label{QuadraticVariety}
\ee
so that the vectors $(b_1, \ldots, b_{d^2})$ lie on the surface of an ellipsoid.

Putting these ingredients together with Eq.~(\ref{traceRhoE}), we
have the following parameterization of valid probability vectors $\|p\drangle$:
\be p(k)=\frac{1}{d}\sum_{ij}c_{ijk}b_i b_j\;.
\label{Shape}
\ee Here the $c_{ijk}$ are the triple-product constants defined in
Eq.~(\ref{TripleSec}) and the $b_i$ satisfy the constraint
(\ref{QuadraticVariety}).
%If one prefers, one could make use of the structure constants %$\alpha_{ijk}$ themselves and alternately write
%\be
%p(k)=\frac{1}{d+1}\sum_{ij}b_i b_j\alpha_{ijk}+\frac{1}{d(d+1)}\;.
%\ee

Finally, let us note what the Hilbert-Schmidt inner product of two quantum
states looks like in SIC terms.  If a quantum state $\rho$ is mapped to
$\|p\drangle$ via a SIC, and a quantum state $\sigma$ is mapped to
$\|q\drangle$, then \bea \tr\rho\sigma &=& d(d+1)\sum_i p(i)q(i)-1
\nonumber\\ &=& d(d+1)\dlangle p\|q\drangle - 1\;.
\label{Bibjangles}
\eea Notice a particular consequence of this: Since $\tr\rho\sigma\ge0$, the
distributions associated with distinct quantum states can never be too
nonoverlapping: \be \dlangle p\|q\drangle \ge \frac{1}{d(d+1)}\;.  \ee

With this development we have given a broad outline of the shape of
quantum-state space in SIC terms.  We do this because that shape is our target.
Particularly, we are obliged to answer the following question: If one takes the
view that quantum states are {\it nothing more\/} than probability
distributions with the restrictions (\ref{Shape}) and (\ref{QuadraticVariety}),
what could motivate that restriction?  That is, what could motivate it
{\it other than\/} knowing the usual formalism for quantum mechanics?  The
answer has to do with rewriting the Born Rule in terms of SICs, which we will
do in Section \ref{KnuckleFinger}.

\subsection{Aside on Unitarity}

Let us take a moment to move beyond statics and rewrite quantum dynamics in SIC
terms: We do this because the result will have a striking
resemblance to the Born Rule itself, once developed in the next section.

Suppose we start with a density operator $\rho$ and let it evolve under unitary
time evolution to a new density operator $\sigma=U\rho U^\dagger$.  If $\rho$
has a representation $p(i)$ with respect to a certain given SIC, $\sigma$ will
have a SIC representation as well---let us call it $q(j)$.  We use the different
index $j$ (contrasting with $i$) to help indicate that we are talking about the
quantum system at a later time than the original.

What is the form of the mapping that takes $\|p\drangle$ to $\|q\drangle$?  It
is simple enough to find with the help of Eqs.~(\ref{Mustard0})
and~(\ref{FierceUrgencyOfNow}):
\bea
q(j)&=&\frac{1}{d}\tr\sigma\Pi_j \nonumber\\ &=&\frac{1}{d}\sum_i
\left((d+1)p(i)-\frac{1}{d}\right)\tr\!\Big(U\Pi_iU^\dagger\Pi_j\Big)\;. \nonumber\\ \;
\eea
If we now define
\be
r_{\rm \scriptscriptstyle U}(j|i)=\frac{1}{d}\tr\!\Big(U\Pi_iU^\dagger\Pi_j\Big)
\ee
and remember, e.g., Eq.~(\ref{psipsi}), we have that
\be
0\le r_{\rm \scriptscriptstyle U}(j|i)\le 1
\ee
and
\be \sum_j r_{\rm \scriptscriptstyle U}(j|i) = 1 \quad
\forall i \qquad\mbox{and}\qquad \sum_i r_{\rm \scriptscriptstyle U}(j|i) = 1 \quad \forall j\;.
\ee
In other words, the $d^2\times d^2$ matrix $[\,r_{\rm \scriptscriptstyle U}(j|i)\,]$ is a doubly stochastic
matrix \cite{Horn85}.

Most importantly, one has
\be
q(j)=(d+1)\sum_{i=1}^{d^2} p(i) r_{\rm \scriptscriptstyle U}(j|i) - \frac{1}{d}\;.
\label{NoAccident}
\ee
Without the $(d+1)$ factor and the $\frac1d$ term, this equation would
represent classical stochastic evolution. Unitary time evolution in a SIC
representation is thus formally close to classical stochastic
evolution. As we shall shortly see, this teaches us something about unitarity
and its connection to the Born Rule itself.

\section{Expressing the Born Rule in Terms of SICs}
\label{KnuckleFinger}

\begin{figure}
\includegraphics[height=3.3in]{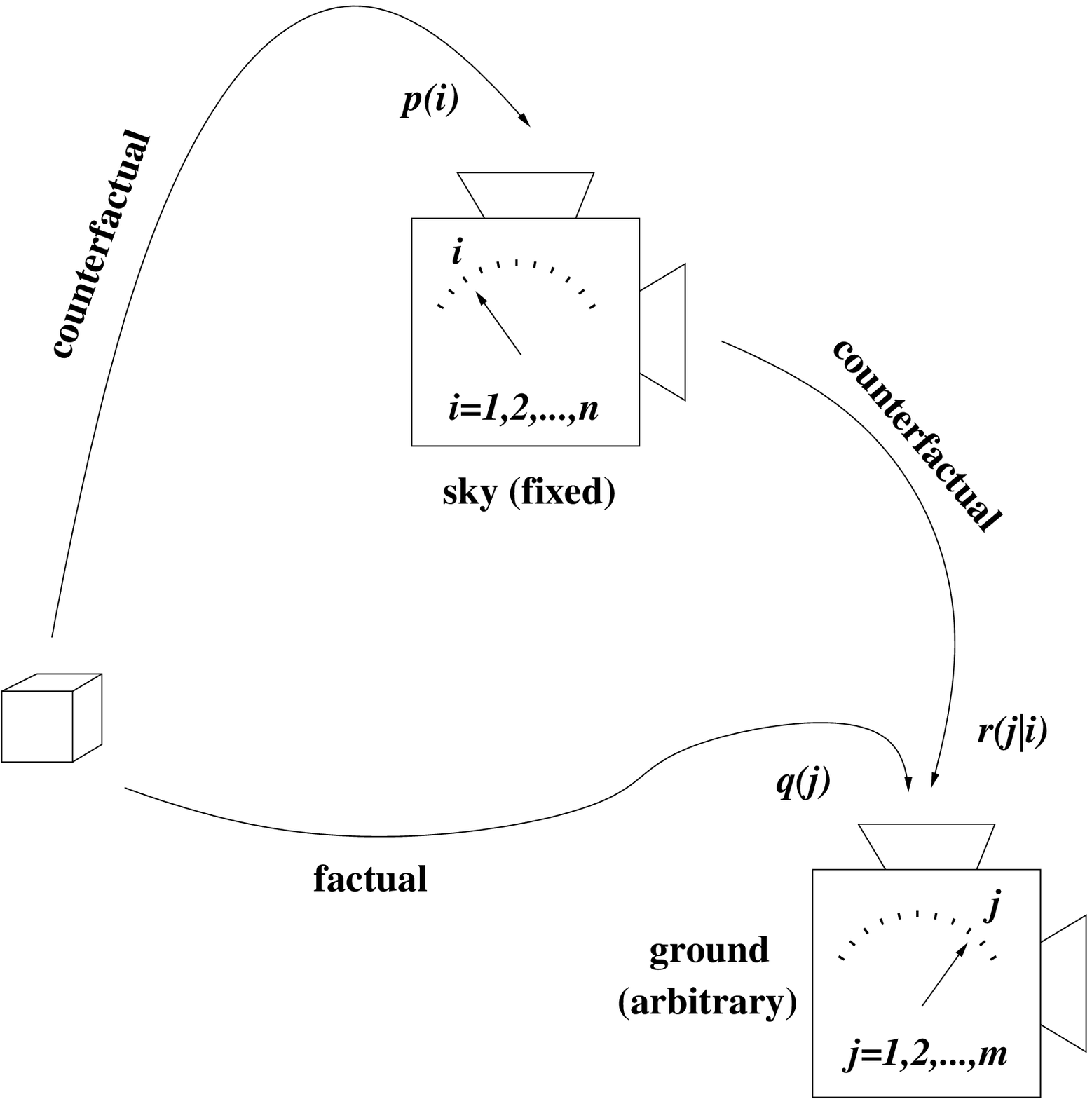}
\bigskip\caption[The basic conceptual apparatus of this paper.]{The
  diagram above expresses the basic conceptual apparatus of this
  paper.  The measurement on the ground, with outcomes $j=1,\ldots,m$,
  is some potential measurement that could be performed in the
  laboratory---i.e., one that could be factualized.  The measurement
  in the sky, on the other hand, with outcomes $i=1,\ldots,n$, is a
  fixed measurement one can contemplate independently.  The
  probability distributions $p(i)$ and $r(j|i)$ represent how an agent
  would gamble if a conditional lottery based on the measurement in
  the sky were operative.  The probability distribution $q(j)$
  represents instead how the agent would gamble on outcomes of the
  ground measurement if the measurement in the sky and the associated
  conditional lottery were nullified---i.e., they were to never take
  place at all.  In the quantum case, the measurement in the sky is a
  SIC with $n=d^2$ outcomes; the measurement on the ground is any
  POVM.  In pure Bayesian reasoning, there is no necessity that $q(j)$
  be related to $p(i)$ and $r(j|i)$ at all.  In quantum mechanics,
  however, there is a very specific relation:
$$
  q(j)=(d+1)\sum_{i=1}^{d^2} p(i) r(j|i) -
     \frac{1}{d}\sum_{i=1}^{d^2} r(j|i)\;.
$$

This equation contains the sum content of the Born Rule, to which it
is equivalent.}
\end{figure}

In this section we come to the heart of the paper: We rewrite the Born Rule in
terms of SICs.  It is easy enough; we just use the expansion in
Eq.~(\ref{FierceUrgencyOfNow}).  Let us first do it for an {\it arbitrary\/}
von Neumann measurement---that is, any measurement specified by a set of rank-1
projection operators $P_j=|j\rangle\langle j|$, $j=1,\ldots,d$.  Expressing the
Born Rule the usual way, we obtain these probabilities for the measurement
outcomes: \be q(j)=\tr\rho P_j\,\;.  \ee Then, by defining \be r(j|i)=\tr \Pi_i
P_j\;,
\label{Amoebus}
\ee
one sees that the Born Rule becomes
\be
q(j)=(d+1)\sum_{i=1}^{d^2} p(i) r(j|i) - 1\;.
\label{PrettyBoyFloyd}
\ee

Let us take a moment to seek out a good interpretation of this equation. It
should be viewed as a direct expression of the considerations laid out in
Section II.  For imagine that before performing the $P_j$ measurement---we will
call it the ``measurement on the ground''---we were to perform a SIC
measurement $\Pi_i$.  We will call the latter the ``measurement in the sky.''

Starting with an initial quantum state $\rho$, we would assign a probability
distribution $p(i)$ to the outcomes of the SIC measurement. In order to be able
to say something about probabilities conditional on a particular outcome of the
SIC measurement, we need to specify the post-measurement quantum state for that
outcome. Here we will adopt the standard L\"uders Rule \cite{Busch95,Busch09}, that $\rho$ transforms to $\Pi_i$ when outcome
$i$ occurs. The conditional probability for getting $j$ in the
subsequent von Neumann measurement on the ground, consequent upon $i$, is then
precisely $r(j|i)$ as defined in Eq.~(\ref{Amoebus}).  With these assignments,
Dutch-book coherence demands an assignment $s(j)$ for the outcomes on the
ground that satisfies \be s(j)=\sum_{i=1}^{d^2} p(i) r(j|i)\;, \ee i.e., a
probability that comes about via the Law of Total Probability, Eq.~(\ref{TotalProb}).

But now imagine the measurement in the sky nullified---i.e., imagine it does
not occur after all---and that the quantum system goes directly to the
measurement device on the ground.  Quantum mechanics tells us to make the
probability assignment $q(j)$ given in Eq.~(\ref{PrettyBoyFloyd}) instead.  So,
\be q(j)=(d+1)s(j)-1 \;.
\label{MorningAfter}
\ee That $q(j)\ne s(j)$ holds, regardless of the assignment of
$s(j)$, is a formal expression of the idea that the ``unperformed SIC
had no outcomes,'' as explained in Sec.~\ref{OhBoy}. But
Eq.~(\ref{MorningAfter}) tells us still more detailed information than this. It
expresses a kind of ``empirically extended coherence''---not implied by Dutch-book coherence
alone, but formally similar to the kind of relation one gets from Dutch-book
coherence. It contains a surprising amount of information about the structure
of quantum mechanics.

To support this, let us try to glean some insight from
Eq.~(\ref{MorningAfter}).  The most obvious thing one can note is that
$\|s\drangle$ cannot be too sharp a probability distribution.  For otherwise
$q(j)$ will violate the bounds $0\le q(j)\le 1$ set by Dutch-book coherence.
Particularly, \be \frac{1}{d+1}\le s(j) \le \frac{2}{d+1}\;.  \ee This in turn
will have implications for the range of values possible for $p(i)$ and
$r(j|i)$.  Indeed if either of these distributions become too sharp (in the
latter case, for too many values of $i$), again the bounds will be violated.
This suggests that an essential
part of quantum-state space structure, as expressed by its extreme points
satisfying Eqs.~(\ref{PurePurity}) and (\ref{MonkeyFun}), arises from the very
requirement that $q(j)$ be a proper probability distribution.  In the next
Section, we will explore this question in greater depth.

First though, we must note the most general form of the Born Rule, when the
measurement on the ground is not restricted to being of the simple von Neumann
variety.  So, let \be q(j)=\tr \rho F_j\ee and \be r(j|i)=\tr \Pi_i
F_j \label{trPiiFj}\ee for some general POVM $\{F_j\}$ on the ground, with any number of
outcomes, $j=1,\ldots,m$.  Then the Born Rule becomes \be
q(j)=(d+1)\sum_{i=1}^{d^2} p(i) r(j|i) - \frac{1}{d}\sum_{i=1}^{d^2} r(j|i)\;.
\label{PrettyBoyBabyThisBeLloyd}
\ee As stated, this is the most general form of the Quantum Law of Total
Probability. It has two terms, a term comprising the classical Law of
Total Probability, and a term dependent only upon the sum of the conditional
probabilities.

When the measurement on the ground is itself another
SIC (any SIC) it reduces to \be q(j)=(d+1)\sum_{i=1}^{d^2} p(i) r(j|i) -
\frac{1}{d}\;.  \ee Notice the formal resemblance between this and
Eq.~(\ref{NoAccident}) expressing unitary time evolution.

\subsection{Why ``Empirically Extended Coherence'' Instead of Objective Quantum States?}

What we are suggesting is that perhaps Eq.~(\ref{PrettyBoyBabyThisBeLloyd})
should be taken as one of the basic axioms of quantum theory, since it provides
a particularly clear way of thinking of the Born Rule as an addition to
Dutch-book coherence. This addition is empirically based and gives extra
normative rules, beyond the standard rules of probability theory, to guide the
agent's behavior when he interacts with the physical world.

But, one may well ask, what is our problem with the
standard way of expressing the Born Rule in the first place?  How is
introducing an addition to Dutch-book coherence conceptually any more palatable
than introducing objective quantum states or objective probability
distributions?  For, if the program is successful, then the demand that $q(j)$
be a proper probability distribution will place necessary restrictions on
$p(i)$ and $r(j|i)$.  This---a skeptic would say---is {\it the\/} very sign
that one is dealing with objective (or agent-independent) probabilities in the
first place.  Why would a personalist Bayesian accept {\it any\/} a priori
restrictions on his probability assignments?  And particularly, restrictions
supposedly of empirical origin?

The reply is this.  It is true that through an axiom like
Eq.~(\ref{PrettyBoyBabyThisBeLloyd}) one gets a restriction on the ranges of
the various probabilities one can contemplate holding.  But that restriction in
no way diminishes the functional role of prior beliefs in the makings of an
agent's particular assignments $p(i)$ and $r(j|i)$.  That is, this addition to
Dutch-book coherence preserves the points expressed in the quote by Keynes in Section II in a way that
objective chance cannot.

Take the usual notion of objective chance, as given operational meaning through
David Lewis' ``Principal Principle'' \cite{Lewis86a,Lewis86b}.  If an
event $A$ has objective chance $\mbox{ch}(A)=x$, then the subjective,
personalist probability an agent (any agent) should ascribe to $A$ on the
condition of knowing the chance proposition is \be
\mbox{Prob}\Big(A\;\Big|\;\mbox{``ch}(A)=x\mbox{''} \wedge E\Big)=x \ee where
$E$ is any ``admissible'' proposition.  There is some debate about what
precisely constitutes an admissible proposition, but an example of a
proposition universally accepted to be compatible in spite of these
interpretive details is this: \bea E &=& \mbox{``All my experience causes me to
  believe $A$} \nonumber\\ && \mbox{ with probability 75\%.''}\nonumber \eea
That is, upon knowing an objective chance, all prior beliefs should be
overridden. Regardless of the agent's firmly held belief about $A$, that belief
becomes irrelevant once he is apprised of the objective chance.

When it comes to quantum mechanics, philosophers of science who find something
digestible in Lewis' idea, often view the Born Rule itself as a healthy
serving of Principal Principle.  Only, it has the quantum state $\rho$ filling
the role of chance.  That is, for any agent contemplating performing a
measurement $\{P_j\}$, his subjective, personal probabilities for the outcomes
$j$ should condition on knowledge of the quantum state just as one conditions
with the Principal Principle: \be \mbox{Prob}\big(j\;\big|\;\rho \wedge
E\big)=\tr\rho P_j\;, \ee where $E$ is any ``admissable'' proposition.  Beliefs
are beliefs, but quantum states are something else: They are the facts of
nature that power a quantum version of the Principal Principle. In other words,
in this context one has conceptually
\be
\rho \quad \longrightarrow \quad \mbox{``ch}(j)=\tr\rho P_j\mbox{''}\;.
\ee

But the Quantum-Bayesian view cannot sanction this.  For, the essential point
for a Quantum Bayesian is that there is no such thing as {\it the\/} quantum
state.  There are potentially as many states for a given quantum system as
there are agents.  And that point is not diminished by accepting the addition
to Dutch-book coherence described in this paper.  Indeed, it is just as with standard
(nonquantum) probabilities, where their subjectivity is not diminished by
normatively satisfying standard Dutch-book coherence.

The most telling reason for this arises directly from quantum statistical
practice.  The way one comes to a quantum-state assignment is ineliminably
dependent on one's priors \cite{Fuchs02,Caves07,Fuchs09}.  Quantum states are
not god-given, but have to be fought for via measurement, updating,
calibration, computation, and any number of related activities.  The only
place quantum states are ``given'' outright---that is to say, the model on
which much of the notion of an objective quantum state arises from in the first
place---is in a textbook homework problem.  For instance, a textbook exercise
might read, ``Assume a hydrogen atom in its ground state. Calculate \ldots.''
But outside the textbook it is not difficult to come up with examples where two
agents looking at the same data, differing only in their prior beliefs, will
asymptotically update to distinct (even orthogonal) pure quantum-state
assignments for the same system \cite{Fuchs09}.\footnote{Here is a simple if
  contrived example. Consider a two-qubit system for which two agents have distinct
  quantum-state assignments $\rho_+$ and $\rho_-$, defined by
$\rho_\pm=\frac12(|0\rangle\langle0|^{\otimes2}+|\pm\rangle\langle\pm|^{\otimes2})$
where $|\pm\rangle=2^{-1/2}(|0\rangle\pm|1\rangle)$. These state assignments are ``compatible'' in several of the senses of \cite{Brun02,Caves02d}, yet suppose the first qubit is
measured in the basis $\{|0\rangle,|1\rangle\}$ and outcome 1 is found. The two
agents' post-measurement states for the second qubit are $|+\rangle$ and $|-\rangle$,
respectively. See \cite{Fuchs09} for a more thorough discussion.\label{NoConvergence}}   Thus the basis for one's
particular quantum-state assignment is always {\it outside\/} the formal
apparatus of quantum mechanics.\footnote{Nor, does it help to repeat over and
  over, as one commonly hears coming from the philosophy-of-physics community,
  ``quantum probabilities are specified by physical law.''  The simple reply
  is, ``No, they are not.''  The phrase has no meaning once one has taken on
  board that quantum states are born in probabilistic considerations, rather
  than being the parents of them, as laboratory practice clearly shows
  \cite{Paris04,Kaznady09}.}

This is the key difference between the set of ideas being developed here and
the position of the objectivists: added relations for probabilities, yes, but no
one of those probabilities can be objective in the sense of being any less a pure
function of the agent.  A way to put it more prosaically is that these
normative considerations may narrow the agent from the full probability simplex
to the set of quantum states, but beyond that, the formal apparatus of quantum
theory gives him no guidance on which quantum state he should choose.  Instead,
the role of a normative reading of the Born Rule is as it is with usual Dutch
book. Here is the way L. J. Savage put it rather eloquently
\cite[p.~57]{Savage54}.  \bq According to the personalistic view, the role of
the mathematical theory of probability is to enable the person using it to
detect inconsistencies in his own real or envisaged behavior.  It is also
understood that, having detected an inconsistency, he will remove it.  An
inconsistency is typically removable in many different ways, among which the
theory gives no guidance for choosing.  \eq If an agent does not satisfy
Eq.~(\ref{PrettyBoyBabyThisBeLloyd}) with his personal probability assignments,
then he is not recognizing properly the change of conditions (or perhaps we
could say `context'\footnote{We add this alternative formulation so as to place
  the discussion within the context of various other analyses of the idea of
  `contextuality'
  \cite{Mermin93,Grangier02,Grangier05,Appleby05c,Spekkens05,Spekkens08,Ferrie08,Ferrie09}.}) that a
potential SIC measurement would bring about.  The theory gives no guidance for
which of his probabilities should be adjusted or how, but it does say that they
must be adjusted or ``undesirable consequences'' will become unavoidable.

Expanding on this point, Bernardo and Smith put it this way in
\cite[p.~4]{Bernardo94}: \bq Bayesian Statistics offers a rationalist
theory of personalistic beliefs in contexts of uncertainty, with the central
aim of characterising how an individual should act in order to avoid certain
kinds of undesirable behavioural inconsistencies.  \ldots\ The goal, in effect,
is to establish rules and procedures for individuals concerned with disciplined
uncertainty accounting.  The theory is not descriptive, in the sense of
claiming to model actual behaviour.  Rather, it is prescriptive, in the sense
of saying `if you wish to avoid the possibility of these undesirable
consequences you must act in the following way.'  \eq So much, indeed, we
imagine for the {\it full\/} formal structure of quantum mechanics (including
dynamics, tensor-product structure, etc.)---that it is all or nearly all an
addition to Dutch-book coherence.  And specifying those ``undesirable
consequences'' in terms independent of the present considerations is a
significant part of the project of specifying the ontology underlying the
Quantum-Bayesian position. But that is a goal we have to leave for future work.
Let us now explore the consequences of adopting
Eq.~(\ref{PrettyBoyBabyThisBeLloyd}) as a basic statement, acting as if we do
not yet know the underlying Hilbert-space structure that gave rise to it.

\section{Deriving Quantum-State Space from ``Empirically Extended Coherence''}

\label{Heimlich}

Let us see how far we can go toward deriving various general features of
quantum-state space from the conceptual apparatus portrayed in Figure 2.
Remember that we are representing quantum states by probability vectors
$\|p\drangle$ lying in the probability simplex $\Delta_{d^2}$. The set of
pure states is given by the solutions of either Eq.~(\ref{MegaMorph}) or
Eqs.~(\ref{PurePurity}) and (\ref{MonkeyFun}), and can thus be thought of as an
algebraic variety within $\Delta_{d^2}$ \cite{Sullivant}. The set of all
quantum states, pure or mixed, is the convex hull of the set of pure states,
i.e., the set of all convex combinations of vectors $\|p\drangle$ representing
pure states. We want to explore how much of this structure can be recovered
from the considerations summarized in Figure 2. We will have to add at least
three other assumptions on the nature of quantum measurement, but at first, let
us try to forget as much about quantum mechanics as we can.

Namely, start with Figure 2 but forget about quantum mechanics and forget about
SICs.  Simply visualize an imaginary experiment in the sky $S$, supplemented
with various real experiments we might perform on the ground $G$.  We postulate
that the probabilities we should ascribe for the outcomes of $G$, are
determined by the probabilities we would ascribe to the imaginary outcomes in
the sky and the conditional probabilities for the outcomes of $G$ consequent
upon them, were the measurement in the sky factualized.  Particularly we take
Eq.~(\ref{PrettyBoyBabyThisBeLloyd}) as a postulate: \be
q(j)=(d+1)\sum_{i=1}^{d^2} p(i) r(j|i) - \frac{1}{d}\sum_{i=1}^{d^2} r(j|i)\;.
\label{WaitingForBread}
\ee
We call this postulate the \textbf{urgleichung} (German for ``primal equation'') to emphasize its
primary status to all our thinking.  As before, $p(i)$ represents the probabilities in the sky and
$q(j)$ represents the probabilities on the ground.  The index $i$ is assumed to
range from 1 to $d^2$, for some fixed natural number $d$.  The range of $j$
will not be fixed, but in any case considered will be denoted as running from 1
to $m$.  (For example, for some cases $m$ might be $d^2$, for some cases it
might be $d$, but it need be neither and may be something else entirely---it
will depend upon which experiment we are talking about for $G$.)  We write
$r(j|i)$ to represent the conditional probability for obtaining $j$ on the
ground, given that the experiment in the sky was actually performed and
resulted in outcome $i$.  When we want to suppress
components, we will write vectors $\|p\drangle$ and $\|q\drangle$, and write
$R$ for the matrix with entries $r(j|i)$. By definition, $R$ is a stochastic
matrix, i.e., $\sum_j r(j|i)=1$, but not necessarily a doubly stochastic
matrix, i.e., $\sum_i r(j|i)=1$ does not necessarily hold
\cite[pp.~526--528]{Horn85}.

One of the main features we will require, of course, is that calculated by Eq.~(\ref{WaitingForBread}), $\|q\drangle$ must satisfy $0\le q(j)\le 1$ for all $j$.  Thus, let us also honor the special inequality
\be
0\;\le\; (d+1)\sum_{i=1}^{d^2} p(i) r(j|i) - \frac{1}{d}\sum_{i=1}^{d^2} r(j|i)\;\le\; 1
\label{TheTrueUr}
\ee
with a name:  the {\it fundamental inequality}.

To proceed, let us define two sets $\mathcal P$ and $\mathcal R$, the first
consisting of priors for the sky $\|p\drangle$, and the second consisting of
stochastic matrices $R$.\footnote{The matrices $R$ could also be regarded as
  part of the agent's prior, but since in this paper we keep $R$ fixed once the
  measurement on the ground is fixed, we reserve the term ``prior'' for members
  of the set ${\mathcal P}$.}  We shall sometimes call $\mathcal P$ our {\it
  state space}, and its elements {\it states}.  We will say that $\mathcal P$
and $\mathcal R$ are {\it consistent\/} (with respect to the fundamental
inequality) if 1) for any fixed $R\in \mathcal R$, there is no
$\|p\drangle\in\mathcal P$ that does not satisfy the fundamental inequality, and
2) for any fixed $\|p\drangle\in\mathcal P$, there is no $R\in \mathcal R$ that
does not satisfy the fundamental inequality.  With respect to consistent sets
$\mathcal P$ and $\mathcal R$, for convenient terminology, we call a general
$\|p\drangle\in\Delta_{d^2}$ {\it valid\/} if it is within the state space
$\mathcal P$; if it is not within $\mathcal P$, we call it invalid.

What we want to pin down are the properties of $\mathcal P$ and $\mathcal R$
under the assumption that they are {\it maximal}.  By this we mean that for any
$\|p^\prime\drangle\notin\mathcal P$, if we were to attempt to create a new
state space ${\mathcal P}^\prime$ by adding $\|p^\prime\drangle$ to the
original $\mathcal P$, ${\mathcal P}^\prime$ and $\mathcal R$ would not be
consistent.  Similarly, if we were to attempt to add a new point $R^\prime$ to
$\mathcal R$.  In other words, when $\mathcal P$ and $\mathcal R$ are maximal,
they are full up with respect to any further additions.  In summary,
\begin{enumerate}
\item
$\mathcal P$ and $\mathcal R$ are said to be {\it consistent\/} if all pairs
  $\big(\|p\drangle,R\big) \in {\mathcal P}\times {\mathcal R}$ obey the fundamental inequality.
\item
$\mathcal P$ and $\mathcal R$ are said to be {\it maximal\/} whenever
  ${\mathcal P}^\prime\supseteq {\mathcal P}$ and ${\mathcal R}^\prime
  \supseteq {\mathcal R}$ imply ${\mathcal P}^\prime={\mathcal P}$ and
  ${\mathcal R}^\prime={\mathcal R}$ for any consistent ${\mathcal P}^\prime$
  and ${\mathcal R}^\prime$.
\end{enumerate}
The key idea behind the demand for maximality is that we want the urgleichung to be as least exclusionary as possible in limiting an agent's probability assignments. There is, of course, no guarantee without further assumptions there will be a
unique maximal $\mathcal P$ and $\mathcal R$ consistent with the fundamental
inequality, or even whether there will be a unique set of them up to
rotations or other kinds of transformations, but we can certainly say some things.

One important result follows immediately: If $\mathcal P$ and $\mathcal R$ are
consistent and maximal, both sets must be convex.  For instance, if
$\|p\drangle$ and $\|p^\prime\drangle$ satisfy (\ref{TheTrueUr}) for all $R\in
\mathcal R$ it is clear that, for any $x\in [0,1]$,
$\|p^{\prime\prime}\drangle=x \|p\drangle + (1-x) \|p^\prime\drangle$ will as
well.  Thus, if $\|p^{\prime\prime}\drangle$ were not in $\mathcal P$, the set would
not have been maximal to begin with.\footnote{It is important to recognize that
  the considerations leading to the convexity of the state space here are
  distinct from the arguments one finds in the ``convex sets'' and
  ``operational'' approaches to quantum theory. See for instance
  \cite{Holevo82,Busch95} and more recently the BBLW school starting in
  \cite{Barnum06} (and several publications thereafter), as well as the
  work of Hardy \cite{Hardy01}.  There the emphasis is on the idea that a state
  of ignorance about a finer preparation is a preparation itself.  The present
  argument even differs from some of our own earlier Bayesian
  considerations (where care was taken {\it not\/} to view `preparation' as an
  objective matter of fact, independent of prior beliefs, as talk of
  preparation would seem to imply) \cite{Fuchs02,Schack04}.  Here instead, the
  emphasis is on the closure of the fundamental inequality, i.e.,
  maximal $\mathcal P$ and $\mathcal R$.}
Furthermore, maximality and the boundedness of Eq.~(\ref{TheTrueUr}) ensures that $\mathcal P$ and $\mathcal R$ are closed sets, thus convex sets with extreme points \cite{Appleby11}.

Now, is there any obvious connection between $\mathcal P$ and $\mathcal R$?
Let us make the  innocuous assumption that one can be completely
ignorant of the outcomes in the sky:
\begin{assump}{\rm $\!\!$:}
\be
\|p\drangle=\left(\frac{1}{d^2},\frac{1}{d^2},\ldots,\frac{1}{d^2}\right)^{\!\rm
  T}\in {\mathcal P}\;.
\ee
\end{assump}

Certainly for any real-world experiment, one can
always be maximally ignorant of which of its outcomes will occur! Suppose now
that the experiment in the sky really is performed as well as the experiment on
the ground. Before either experiment, the agent is ignorant of both the outcome $i$
in the sky and the outcome $j$ on the ground. Using Bayes' rule, he can
find the conditional probability for $i$ given $j$, which has the form of a
posterior probability, \be
\mbox{Prob}(i|j)=\frac{r(j|i)}{\sum_k r(j|k)}\;.
\label{NoseHair}
\ee
Let us now make a less innocuous assumption:
\begin{assump}{\rm $\!\!$:}
{\rm Principle of Reciprocity:\ Posteriors from Maximal Ignorance Are Priors.}  For any $R\in\mathcal
R$, a posterior probability $\mbox{\rm Prob}(i|j)$ as in Eq.~(\ref{NoseHair})
is a valid prior $p(i)$ for the outcomes of the measurement in the
sky. Moreover, for each valid $p(i)$, there exists some $R\in\mathcal
R$ and some $j$ such that $p(i)=\mbox{Prob}(i|j)$ as in Eq.~(\ref{NoseHair}).
\label{Reciprocity}
\end{assump}
Quantum mechanics certainly has this property.  For, suppose a completely mixed
state for our quantum system and a POVM ${\mathcal G}=\{G_j\}$ measured on the
ground.  Upon noting an outcome $j$ on the ground, the agent will use
Eqs.~(\ref{trPiiFj}) and (\ref{NoseHair}) to infer \be
\mbox{Prob}(i|j)=\frac{\tr\Pi_i G_j}{d\,\tr G_j}\;.  \ee Defining \be
\rho_j=\frac{G_j}{\tr G_j}\;, \ee this says that \be
\mbox{Prob}(i|j)=\frac{1}{d}\,\tr\rho_j\Pi_i\;.  \ee In other words,
$\mbox{Prob}(i|j)$ is itself a SIC-representation of a quantum state.
Moreover, $\rho_j$ can be any quantum state whatsoever, simply by adjusting
which POVM $\mathcal G$ is under consideration.

\subsection{Basis Distributions}

\label{BengalTiger}

Since we are free to contemplate any measurement on the ground, let us consider
the case where the ground measurement is set to be the same as that of the sky.
We will denote $r(j|i)$ by $r_{\rm \scriptscriptstyle S}(j|i)$ in this special
case. Remembering that the probabilities on the ground, $q(j)$, refer to the case that the
measurement in the sky remains counterfactual, we must then have that $p(j)=q(j)$
for any valid $\|p\drangle$, or, using the urgleichung,
\be p(j)=(d+1)\sum_i p(i)r_{\rm \scriptscriptstyle S}(j|i) - \frac{1}{d} \sum_i
r_{\rm \scriptscriptstyle S}(j|i)\;.  \label{groundEqualsSky} \ee Take the case
where $p(i)=\frac{1}{d^2}$ specifically. Substituting for $p(i)$ in
Eq.~(\ref{groundEqualsSky}), we find that the $r_{\rm \scriptscriptstyle
  S}(j|i)$ must satisfy \be \sum_i r_{\rm \scriptscriptstyle S}(j|i)=1\;.  \ee
Therefore, when going back to more general priors $\|p\drangle$, one has in
fact the simpler relation \be p(j)=(d+1)\sum_i p(i)r_{\rm \scriptscriptstyle
  S}(j|i) - \frac{1}{d}\;.
\label{FemtoFoot}
\ee
Introducing an appropriately sized matrix $M$ of the form
\be
M=\left(
  \begin{array}{cccc}
    (d+1)-\frac{1}{d} & -\frac{1}{d} & \cdots & -\frac{1}{d} \\
    -\frac{1}{d} & (d+1)-\frac{1}{d} & \cdots &-\frac{1}{d} \\
    \vdots & & \ddots & \vdots \\
    -\frac{1}{d} & -\frac{1}{d} & \cdots & (d+1)-\frac{1}{d} \\
  \end{array}
\right)
\ee
we can rewrite Eq.~(\ref{FemtoFoot}) in vector form,
\be
M R_{\rm \scriptscriptstyle S} \|p\drangle=\|p\drangle\;.
\label{HerbertFoot}
\ee

At this point, we pause for a minor assumption on our state space:
\begin{assump}
The elements $\|p\drangle\in{\mathcal P}$ span the full simplex $\Delta_{d^2}$.
\label{BellyRoar}
\end{assump}
This is a very natural assumption: For if ${\mathcal P}$ did not span the
simplex, one would be justified in simply using a smaller simplex for all
considerations.

With Assumption \ref{BellyRoar}, the only way Eq.~(\ref{HerbertFoot}) can be satisfied is if
\be
M R_{\rm \scriptscriptstyle S} = I\;.
\ee
Since $M$ is a circulant matrix, its inverse is a circulant matrix as well, and one can easily work out that,
\be
r_{\rm \scriptscriptstyle S}(j|i)=\frac{1}{d+1}\left(\delta_{ij}+\frac{1}{d}\right)\;.
\ee
It follows by the Principle of Reciprocity (our Assumption \ref{Reciprocity})
then that among the distributions in $\mathcal P$, along with the uniform distribution, there are at least $d^2$ other ones, namely:
\be
\|e_k\drangle= \left(\, \frac{1}{d(d+1)}, \ldots, \frac{1}{d}, \ldots, \frac{1}{d(d+1)}\,\right)^{\!\rm T}\;,
\ee
with a $\frac{1}{d}$ in the $k^{\rm th}$ slot and $\frac{1}{d(d+1)}$ in all other slots.  We shall call these $d^2$ special distributions, appropriately enough, the {\it basis distributions}.

Notice that, in the special case of quantum mechanics, the basis distributions are just the SIC states themselves, now justified in a more general setting.
Also, like the SIC states, we will have
\be
\sum_i e_k(i)^2=\frac{2}{d(d+1)}\quad\forall k\,,
\label{EvaCassidyMoment}
\ee
in accordance with Eq.~(\ref{PurePurity}).

\subsection{A Bloch Sphere}

Consider a class of measurements for the ground that have a property we shall
call {\it in-step unpredictability}, ISU.  The property is this:  Whenever one
assigns a uniform distribution for the measurement in the sky, one also assigns
a uniform distribution for the measurement on the ground. This is meant to
express the idea that the measurement on the ground has no in-built bias with
respect to one's expectations of the sky: Complete ignorance of the outcomes of
one translates into complete ignorance of the outcomes of the other.  (In the
full-blown quantum mechanical setting, this corresponds to a POVM $\{G_j\}$
such that $\tr G_j$ is a constant value---von Neumann measurements
with $d$ outcomes being one special case of this.)

Denote the $r(j|i)$ and corresponding matrix $R$ in this special case by
$r_{\rm \scriptscriptstyle ISU}(j|i)$ and  $R_{\rm \scriptscriptstyle
  ISU}$, respectively, and suppose the measurement being spoken of has $m$ outcomes.  Our demand is that
\be
\frac{1}{m}=\frac{(d+1)}{d^2}\sum_i r_{\rm \scriptscriptstyle ISU}(j|i) - \frac{1}{d} \sum_i r_{\rm \scriptscriptstyle ISU}(j|i)\;.
\ee
To meet this, we must have
\be
\sum_i r_{\rm \scriptscriptstyle ISU}(j|i)=\frac{d^2}{m}\;,
\ee
and the urgleichung becomes
\be
q(j)=(d+1)\sum_i p(i) r_{\rm \scriptscriptstyle ISU}(j|i) - \frac{d}{m}\;.
\label{Lichen}
\ee

Suppose now that a prior $\|s\drangle$ for the sky happens to arise in accordance with Eq.~(\ref{NoseHair}) for one of these ISU measurements.  That is,
\be
s(i)=\frac{r_{\rm\scriptscriptstyle ISU}(j|i)}{\sum_k r_{\rm\scriptscriptstyle ISU}(j|k)}\;,
\label{FlotsamJetsam}
\ee
for some $R_{\rm\scriptscriptstyle ISU}$ and some $j$.
Then Eq.~(\ref{Lichen}) tells us that for any $\|p\drangle\in\mathcal P$, we must have
\be
0\,\le\, \frac{d^2}{m}(d+1)\sum_i p(i) s(i) - \frac{d}{m}\,\le\, 1\;.
\ee
In other words, for any $\|s\drangle$ of our specified variety and any $\|p\drangle\in\mathcal P$, the following constraint must be satisfied
\be
\frac{1}{d(d+1)}\,\le\, \sum_i p(i) s(i) \,\le\, \frac{d+m}{d^2(d+1)}\;.
\ee

Think particularly on the case where $\|s\drangle=\|p\drangle$.  Then we must have
\be
\sum_i p(i)^2 \,\le\, \frac{d+m}{d^2(d+1)}\;.
\label{MommaToldMeNotToCome}
\ee
Note how this compares to Eq.~(\ref{EvaCassidyMoment}).

Now, suppose there are ISU measurements (distinct from simply bringing the sky
measurement down to the ground) that have the basis distributions
$\|e_k\drangle$ as their posteriors in the way of Assumption \ref{Reciprocity}, the Principle of
Reciprocity.  If this is so, then note that according to
Eq.~(\ref{EvaCassidyMoment}) the bound in Eq.~(\ref{MommaToldMeNotToCome})
will be violated unless $m\ge d$.  Moreover, it will not be tight for the basis
states unless $m=d$ precisely.

Thinking of a basis distribution as the prototype of an extreme-point state
(for after all, they give the most predictability possible for the measurement
in the sky), this motivates the next assumption---this one being significantly
stronger than the previous two:
\begin{assump}
Every extreme point $\|p\drangle\in\mathcal P$ arises
in the manner of Eq.~(\ref{FlotsamJetsam}) as the posterior of an
ISU measurement with $m=d$
and achieves equality in Eq.~(\ref{MommaToldMeNotToCome}).
\end{assump}
Thus, for {\it any\/} two extreme points $\|p\drangle$ and $\|s\drangle$, we are assuming
\be
\frac{1}{d(d+1)}\,\le\, \sum_i p(i) s(i) \,\le\, \frac{2}{d(d+1)}\;,
\label{SayHiToYourKnee}
\ee with equality in the right-hand side when
$\|s\drangle=\|p\drangle$.\footnote{By the way, it should be noted that this
  inequality establishes that if $\mathcal P$ at least contains the actual
  quantum-state space, it can contain no more than that.  That is, the full set
  of quantum states is, in fact, a maximal set. For suppose a SIC exists, yet
  $\|s\drangle$ corresponds to some non-positive-semidefinite operator via the
  mapping in Eq.~(\ref{FierceUrgencyOfNow}).  Then there will be some
  $\|p\drangle\in \mathcal P$ corresponding to a pure quantum state such that
  the left-hand side of Eq.~(\ref{SayHiToYourKnee}) is violated.  This follows
  immediately from the definition of positive semi-definiteness and the
  expression for Hilbert-Schmidt inner products in Eq.~(\ref{Bibjangles}).}

Thus, the extreme points of $\mathcal P$ live on a sphere
\be
\sum_i p(i)^2 \,=\, \frac{2}{d(d+1)}\;,
\label{JiggleMeBoog}
\ee Further trivial aspects of quantum-state space follow
immediately from the requirement of Eq.~(\ref{SayHiToYourKnee}) for any two extreme
points.  For instance, since the basis distributions are among the set of
valid states, for any other valid state $\|p\drangle$ no component in it can be
too large.  This follows because \be \dlangle p\|
e_k\drangle=\frac{1}{d(d+1)}+\frac{1}{d+1}p(k)\;.  \ee The right-hand side of
Eq.~(\ref{SayHiToYourKnee}) then requires \be p(k)\le \frac{1}{d}\;.  \ee But,
do we have enough to get to us all the way to Eq.~(\ref{ChocolateMouse}) in
addition to Eq.~(\ref{PurePurity})?  We will analyze aspects of this in the
next subsection.  First however, let us linger a bit over the significance of
the sphere.

What we have postulated in a natural way is that the extreme points of
$\mathcal P$ must live on a $(d^2-1)$-sphere centered at the zero vector.  But
then it comes for free that these extreme points must also live on a
smaller-radius $(d^2-2)$-sphere centered at
\be
\|c\drangle=\left(\frac{1}{d^2}, \frac{1}{d^2}, \ldots,
\frac{1}{d^2}\right)^{\!\rm T}\;.
\label{VondaShepard}
\ee
This is because the $\|p\drangle$ live
on the probability simplex $\Delta_{d^2}$.  For, let
$\|w\drangle=\|p\drangle-\|c\drangle$, where $\|p\drangle$ is any point
satisfying Eq.~(\ref{JiggleMeBoog}).  Then \be r^2=\dlangle w\|
w\drangle=\frac{d-1}{d^2(d+1)}
\label{SherylCrow}
\ee
gives the radius of the lower-dimensional sphere.

The sphere in Eq.~(\ref{SherylCrow}) is actually the more natural sphere for us
to think about, as most of the sphere in Eq.~(\ref{JiggleMeBoog})---all but a
set of measure zero---is thrown away anyway.  In fact, it may legitimately be
considered the higher-dimensional analog of the Bloch sphere from the
Quantum-Bayesian point of view.  Indeed, when $d=2$, we have a $2$-sphere, and
it is isomorphic to the usual Bloch sphere.

It is natural to think of the statement
\be
\sum_i p(i)^2 \,\le\, \frac{2}{d(d+1)} \qquad \mbox{for all $\|p\drangle\in \mathcal P$}
\label{WoohMaBoog!}
\ee in information theoretic terms.  This is because two well-known measures of
the uncertainty associated with a probability assignment---the Renyi and
Dar\'oczy entropies \cite{Aczel75} of order 2---are simple functions of the
left-hand side of it.  Recall the Renyi entropies most generally (defined for
all $\alpha\ge1$) \be R_\alpha(\|p\drangle)=\frac{1}{1-\alpha}\ln\!\left(\sum_i
p(i)^\alpha\right) \ee as well as the Dar\'oczy entropies \be
D_\alpha(\|p\drangle)=\frac{1}{2^{1-\alpha}-1}\left(\sum_i
p(i)^\alpha-1\right)\;.  \ee In the limit $\alpha\rightarrow 1$, these both
converge to the Shannon entropy. The characterization of quantum-state space
appears to provide an application of these entropies for the value $\alpha=2$.

To put it in a slogan \cite{Fuchs01,Caves96}, ``In quantum mechanics,
maximal information is not complete and cannot be completed.''  The sharpest
predictability one can have for the outcomes of a SIC measurement is specified
by Eq.~(\ref{PurePurity}).  This is an old idea, of course, but quantified here
in yet another way. It is related to the basic idea underlying the toy model of
R.~W. Spekkens \cite{Spekkens07}, with its ``knowledge balance principle.'' In that model, which combines local hidden
variables with an ``epistemic constraint'' on an agent's knowledge of the
variables' values, more than twenty well-known quantum information theoretic
phenomena (like no-cloning \cite{Wootters82,Dieks82}, no-broadcasting
\cite{Barnum96}, teleportation \cite{Bennett93}, correlation monogamy
\cite{Coffman00}, ``nonlocality without entanglement'' \cite{Bennett99},
etc.)\ are readily reproduced, at least in a qualitative way.

Despite the toy model's impressive successes, however, we suspect that an information
constraint alone cannot support the more sweeping part of the Quantum Bayesian
program, that ``the possible outcomes cannot correspond to actualities,
existing objectively prior to asking the question,'' i.e., that unperformed
measurements have no outcomes. In ways, there is a world of difference between
the present considerations to do with an addition to Dutch-book coherence and
``epistemic restriction'' approaches.  First, it is hard to see how that line
of thought can get beyond the possibility of an underlying hidden-variable
model (as the toy model illustrates). Second, and more importantly, in the
present approach the Bloch sphere may well express an epistemic constraint---a
constraint on an agent's advised certainty.  But the epistemic constraint is
itself a result of a deeper consideration to do with the coherence between
factual and counterfactual gambles, not a starting point. Furthermore, the
constraint is not expressible in terms of a single information function
anyway; instead it involves pairs of distributions.  We go on to explain this
point.

\subsection{But Only Part of It}

\label{Blubbery}

The state-space implied by Eq.~(\ref{SayHiToYourKnee}) does not lead to the
full sphere in Eq.~(\ref{SherylCrow}). According to the left-hand side of
Eq.~(\ref{SayHiToYourKnee}), when two points
are too far away from each other, at least one of them cannot be in $\cal P$.
We will show this more carefully in the next section: that the extreme
$\|p\drangle\in\cal P$ comprise only part of a sphere.  Of some interest,
however, is that Eq.~(\ref{SherylCrow}) already tells us that we cannot have
the full sphere as well.  For, the radius of the sphere is such that the sphere
extends beyond the boundary of the probability simplex $\Delta_{d^2}$.  Hence,
$\cal P$ is contained within a nontrivial intersection of sphere and simplex.

This is established by a nice argument due to Gabriel Plunk \cite{Plunk02}.
Let us calculate the shortest distance between $\|c\drangle$ and an $n$-flat of
the simplex---an $n$-flat is defined so that it contains only
probability vectors with $n$ vanishing components.  For instance, all
$\|p\drangle$ of the form \be \|p_n\drangle=\Big(p(1), p(2), \ldots, p(d^2-n),
0, 0, \ldots, 0\Big)^{\!\rm T}\;, \ee with $d^2-n$ initial nonvanishing
components and $n$ final vanishing components, form an $n$-flat.  A more general $n$-flat would
have all the vanishing and nonvanishing components interspersed.

What is the minimal distance $D_{\rm min}\Big(\|c\drangle,\|p_n\drangle\Big)$
between the center point and an $n$-flat?  Taking \bea
D^2\Big(\|c\drangle,\|p_n\drangle\Big)&=&\sum_{i=1}^{d^2-n}\!\left(p_n(i)-
\frac{1}{d^2}\right)^{\!2} \nonumber\\ &+&
\sum_{i=d^2-n+1}^{d^2}\!\!\left(0-\frac{1}{d^2}\right)^{\!2} \eea generally,
and recognizing the constraint \be \sum_{i=1}^{d^2-n} p_n(i)=1\;, \ee we can
use the calculus of variations to find
\be
D_{\rm  min}^2\Big(\|c\drangle,\|p_n\drangle\Big)=\frac{n}{d^2(d^2-n)}\;,
\ee
Can there be an $n$ for which
\be
D_{\rm min}^2\Big(\|c\drangle,\|p_n\drangle\Big)<
r^2\;?
\ee
where $r$ is defined by Eq.~(\ref{SherylCrow})?
In other words, can the sphere ever poke outside of the probability
simplex? Just solving the inequality for $n$ gives $n<\frac{1}{2}d(d-1)$.

Thus, when $n<\frac{1}{2}d(d-1)$, the point \be
\|p_{s(n)}\drangle\equiv\left(\frac{1}{d^2-n},\frac{1}{d^2-n},\ldots,
\frac{1}{d^2-n},0,0,\ldots,0\right) \ee on an $n$-flat surface of the simplex
lies within the sphere the extreme points of $\cal P$ inhabit.  Only in
the case of the qubit, $d=2$, does the sphere reside completely within the
simplex---the set is equivalent to the well-known Bloch sphere.

A corollary to Plunk's derivation is that we can put a (weak) bound on the
maximum number of zero components a valid $\|p\drangle$ can contain. To have
$n$ zero components, $\|p\drangle$ must live on an $n$-flat.  But extreme
$\|p\drangle$ are always a distance $D^2_{\rm extreme}=\frac{d-1}{d^2(d+1)}$
from $\|c\drangle$.  So, if $n$ is such that \be D^2_{\rm
  min}\Big(\|c\drangle,\|p_n\drangle\Big)>D^2_{\rm extreme}\;, \ee then
$\|p\drangle$ cannot live on the $n$-flat.  This limits $n$: If
$n>\frac{1}{2}d(d-1)$, then a state cannot live on that $n$-flat.
Thus, for a valid $\|p\drangle$, there is an upper bound to how many zero
components it can have\footnote{Our first inclination was that this is surely a
  weak bound.  But even in quantum mechanics, we know of no better
  bound than this.  This follows from the best bound we are aware of in that
  context, a Hilbert-space bound of Delsarte, Goethels, and Seidel
  \cite{Delsarte75} (which we note can also be proven by elementary Gram matrix
  methods in Hilbert-Schmidt space).  Let $P_i$, $i=1, ..., v$, be a set of
  rank-1 projection operators on an $f$-dimensional Hilbert space ${\mathcal
    H}_f$ such that $\tr P_i P_j = c$, for all $i\ne j$.  Then
$$
v\le \frac{f(1-c)}{1-fc}\;.
$$ To find the maximum number of zero components $\|p\drangle$ can contain, we
just need to ask the question of how many SIC vectors can possibly fit in a
$(d-1)$-dimensional subspace.  Inserting the parameters $f=d-1$ and $c=1/(d+1)$
into this bound, we find $n_{\rm zeros}\le \frac{1}{2}d(d-1)$.  Interestingly,
this bound is saturated when $d=2$ and $d=3$. On the other hand, in dimensions
$d=4$ and $d=5$, D. M. Appleby has checked exhaustively for the known SICs that
never more than $d-1$ of the vectors fit within a $(d-1)$-dimensional subspace
\cite{Appleby08pc}.}: \be n_{\rm zeros}\le \frac{1}{2}d(d-1)\;.
\label{NumZeroes}
\ee However, an alternative and more direct argument for Eq.~(\ref{NumZeroes})
is this---it is a straightforward application of the Schwarz inequality: \bea
1&=&\left(\sum_{\stackrel{\rm nonzero}{\rm \scriptscriptstyle
    terms}}p(i)\right)^{\!2} \nonumber\\ &\le&\Big(d^2-n_{\rm
  zeros}\Big)\!\left(\sum_{\stackrel{\rm nonzero}{\rm \scriptscriptstyle
    terms}}p(i)^2\right) \nonumber\\ &=&\Big(d^2-n_{\rm
  zeros}\Big)\frac{2}{d(d+1)}\;.  \eea Eq.~(\ref{NumZeroes}) follows
immediately.

But this is only the beginning of the trimming of the Bloch sphere: More
drastic restrictions come from the left-hand of the fundamental
inequality.

\subsection{An Underlying `Dimensionality'?}

\label{MorningOfLittleHope}

What else does the inequality in Eq.~(\ref{SayHiToYourKnee}) imply?  Here is at
least one more low hanging fruit. The left side of Eq.~(\ref{SayHiToYourKnee})
signifies that the ``most orthogonal'' two valid distributions $\| p\drangle$
and $\| q\drangle$ can ever be is \be \dlangle p\| q\drangle=\sum_i
p(i)q(i)=\frac{1}{d(d+1)}\;.  \ee Their overlap can never approach zero; they
can never be truly orthogonal. Now, suppose we have a collection of
distributions $\| p_k\drangle$, $k=1,\ldots,n$, all of which live on the
sphere---that is, they individually saturate the right-hand side of
Eq.~(\ref{SayHiToYourKnee}).  We can ask, how large can the number $n$ can be
while maintaining that each of the $\| p_k\drangle$ be maximally orthogonal to
each other.  Another way to put it is, what is the maximum number of ``mutually
maximally distant'' states?

In other words, we would like to satisfy \be \dlangle p_k\|
p_l\drangle=\frac{\delta_{kl}+1}{d(d+1)} \ee for as many values as possible.
It turns out that there is a nontrivial constraint on how large $n$ can be, and
it is nothing other than $n=d$---the same thing one sees in quantum mechanics.

To see this, let us again reference the center of the probability simplex with all our vectors.  Define
\be
\|w_k\drangle=\|p_k\drangle-\|c\drangle\;.
\ee
In these terms, our constraint becomes
\be
\dlangle w_k\| w_l\drangle=\frac{d\delta_{kl}-1}{d^2(d+1)}\;.
\ee
However, notice what this means:  We are asking for a set of vectors whose Gram matrix $G=[\dlangle w_k\| w_l\drangle]$ is an $n\times n$ matrix of the form
\be
G=
\left(
  \begin{array}{ccccc}
    a & b & b & \cdots & b \\
    b & a & b & \cdots & b \\
    \vdots & & & \ddots & \vdots \\
    b & b & b & \cdots & a \\
  \end{array}
\right)
\label{113}
\ee with \be a=\frac{d-1}{d^2(d+1)}\qquad\mbox{and}\qquad
b=\frac{-1}{d^2(d+1)}\;.  \ee By an elementary theorem in linear algebra, a
{\it proposed\/} set of vectors with a {\it proposed\/} Gram matrix $G$ can
exist if and only if $G$ is positive semi-definite \cite[pp.~407--408]{Horn85}.
Moreover, the rank of $G$ represents the number of linearly independent such
vectors. (We write ``proposed'' because if $G$ is not positive semi-definite,
then of course there are no such vectors.)

Since $G$ in Eq.~(\ref{113}) is a circulant matrix, its eigenvalues can be
readily calculated: one takes the value \be
\lambda_0=a+(n-1)b=\frac{d-n}{d^2(d+1)} \ee while all the $n-1$ others are \be
\lambda_k=a-b=\frac{1}{d(d+1)}\;.  \ee To make $G$ positive semi-definite,
then, we must have $n\le d$, with $n=d$ being the maximal value.  At that point
$G$ is a rank-$(d-1)$ matrix, so that only $d-1$ of the $\| w_l\drangle$ are
linearly independent.

On the other hand, all $d$ vectors $\| p_k\drangle=\| w_k\drangle+\| c\drangle$
actually are linearly independent.  To see this, suppose there are numbers
$\alpha_i$ such that $ \sum_i \alpha_i \| p_i\drangle=0\;.  $ Acting from the
left on this equation with $\dlangle c\|$, one obtains \be \sum_i \alpha_i=0\;.
\ee On the other hand, acting on it with $\dlangle p_k\|$, we obtain \bea
0&=&\frac{2}{d(d+1)}\alpha_k +\frac{1}{d(d+1)}\sum_{i\ne k}\alpha_i
\nonumber\\ &=& \frac{1}{d(d+1)}\alpha_k + \frac{1}{d(d+1)}\sum_i\alpha_i
\nonumber\\ &=& \frac{1}{d(d+1)}\alpha_k\;.  \eea So indeed, \be \sum_i
\alpha_i\| p_i\drangle=0\qquad\Longrightarrow\qquad \alpha_k=0 \;\;\forall k\;.
\ee

What this reveals is a significantly smaller ``dimension'' for the valid states
on the surface of the sphere than one might have thought. A priori, one might
have thought that one could get nearly $d^2$ maximally equidistant points on
the sphere, but it is not so---only $d$ instead. This is certainly a suggestive
result, but ``dimension'' at this stage must remain in quotes.  Ultimately one
must see that the Hausdorff dimension of the manifold of valid extreme states
is $2d-2$ (i.e., what it is in quantum theory), and the present result does not
get that far.

\subsection{Summary of the Argument So Far}

\setcounter{assump}{-1}

Let us summarize the assumptions made to this point and summarize their
consequences as well.

\begin{assump}{\rm $\!\!$:}
{\rm The Urgleichung.}  See Figure 2.  Degrees of belief for outcomes
in the sky and degrees of belief for outcomes on the ground ought to be related
by this fundamental equation: \be q(j)=(d+1)\sum_{i=1}^{d^2} p(i) r(j|i) -
\frac{1}{d}\sum_{i=1}^{d^2} r(j|i)\;.
\label{Reprise}
\ee
\end{assump}
From the urgleichung, the fundamental inequality arises by the requirement that $0\le q(j)\le 1$ always.  The sets $\mathcal P$ and $\mathcal R$ are defined to be sets of priors $\|p\drangle$ and stochastic matrices $R$, that are consistent and maximal.

\begin{assump}{\rm $\!\!$:}
{\rm The state of maximal ignorance $\|c\drangle$ in Eq.~(\ref{VondaShepard}) is included in $\mathcal P$.}
\end{assump}

\begin{assump}{\rm $\!\!$:}
{\rm Principle of Reciprocity:\ Posteriors from Maximal Ignorance Are Priors.}  For any $R\in\mathcal R$, a posterior probability consequent upon outcome $j$ of a ground measurement,
\be
\mbox{\rm Prob}(i|j)=\frac{r(j|i)}{\sum_k r(j|k)}
\label{ChickenThigh}
\ee
may be taken as a valid prior $p(i)$ for the outcomes of the measurement in the sky. Moreover, all valid priors $p(i)$ may arise in this way.
\end{assump}

\begin{assump}{\rm $\!\!$:}
{\rm $\mathcal P$ spans the probability simplex $\Delta_{d^2}$.}
\end{assump}

\begin{assump}{\rm $\!\!$:}
{\rm Extreme-Point Preparations.}  The extreme points of the convex set
$\mathcal P$ may all be generated as the posteriors of a suitably chosen ground
measurement for which maximal ignorance of sky outcomes implies maximal
ignorance of ground outcomes.  Moreover, these measurements all have the
minimum number of outcomes consistent with generating the basis distributions
$\|e_k\drangle$ in this way.
\end{assump}

With these four assumptions, we derived that the basis distributions
$\|e_k\drangle$ should be among the valid states $\cal P$.  We derived that for
any $\|p\drangle\in \cal P$, the probabilities are bounded above by $p(k)\le
\frac{1}{d}$.  We derived that the extreme points of the valid $\|p\drangle$
should live on the surface of a sphere that
at times pokes outside the probability simplex. We found a bound, given in
Eq.~(\ref{NumZeroes}),
on the number of zero components of $\|p\drangle$ that is as good as the best known bound
that has been derived using the conventional quantum formalism.
Most particularly we derived that for
any two valid distributions $\|p\drangle$ and $\|s\drangle$ (including the case
where $\|p\drangle=\|s\drangle$), it must hold that \be \frac{1}{d(d+1)}\,\le\,
\sum_i p(i) s(i) \,\le\, \frac{2}{d(d+1)}\;.
\label{PimpStick}
\ee From the latter, it follows that no more than $d$ extreme points
$\|p\drangle$ can ever be mutually maximally distant from each
other. Furthermore we showed that not every flat zeros-bound vector can be a
valid $\|p\drangle$.

These are all hints that our structure might just be isomorphic to
quantum-state space under the assumption that SICs exist.
What really needs to be derived is that the extreme points of
such a convex set correspond to an algebraic variety of the form \be
p(k)=\frac{(d+1)^2}{3d}\sum_{ij} c_{ijk}\,p(i)p(j)-\frac{1}{3d}\;, \ee as given
in Eq.~(\ref{MegaMorph}), with a set of $c_{ijk}$
that can be written in the form of Eq.~(\ref{TripleSec}).
Whether this step can be made without making any further
assumptions, we do not know.  Nor do we have a strong feeling presently of
whether the auxiliary Assumptions 1--4 are the ones best posited for
achieving our goal. The key idea is to supplement Assumption 0 with as little extra
structure as possible for getting all the way to full-blown quantum mechanics.
Much work remains, both at the technical and conceptual level.

\section{Relaxing the Constants and Regaining Them}

\setcounter{resump}{-1}

But what is the origin of the urgleichung in the first place?  In this
Section, which in part follows closely \cite{FuchsSchack2011},
we take a small step toward a deeper understanding of the particular
form our relation takes in Eq.~(\ref{Reprise}).  We do this by initially generalizing away from
Eq.~(\ref{Reprise}) and then testing what it takes to get back to it.

What we mean by this is that we should imagine the more general set-up in
Figure 2, where the number of outcomes for the measurement in the sky is
potentially some more general number $n$ (not initially assumed to be a perfect
square $d^2$).  Furthermore we drop away all traces of the parameter $d$, by
considering a generalized urgleichung with two initially arbitrary
parameters $\alpha$ and $\beta$.  That is to say, for this Section, our
fundamental postulate will be:
\begin{resump}{\rm $\!\!$:} {\rm Generalized Urgleichung}.
For whichever experiment we are talking about for the ground, $q(j)$ should be calculated according to
\be
q(j)=\alpha\sum_{i=1}^n p(i)r(j|i) - \beta \sum_{i=1}^n r(j|i)\;,
\label{BigBoy}
\ee
where $\alpha$ and $\beta$ are fixed nonnegative real numbers.
\end{resump}
Otherwise, all considerations will be the same as they were in the beginning of
Section \ref{Heimlich}.  Particularly, the measurements on the ground can have
any number $m$ of outcomes, where the value $m$ in any individual case will be
set by the details of the measurement under consideration at that time.  Our
goal will be to see what assumptions can be added to this basic scenario so
that the urgleichung in Eq.~(\ref{Reprise}) re-arises in a natural
manner.  That is to say, we would like to see what assumptions can be added to
this recipe so that $\alpha=d+1$, $\beta=\frac{1}{d}$, and $n=d^2$ (for some
$d$) are the end result.

Immediately, one can see that $n$, $\alpha$, and $\beta$ cannot be independent.
This just follows from the requirements that \be \sum_{j=1}^m q(j) = 1, \quad
\sum_{j=1}^m r(j|i) =1 \ \ \forall i, \quad \mbox{and} \quad \sum_{i=1}^n
p(i)=1\;.  \ee Summing both left and right sides of Eq.~(\ref{BigBoy}) over
$j$, one obtains, \be n \beta = \alpha-1\;.
\label{RainyDay}
\ee
Furthermore, since $\beta\ne 0$ is assumed, requiring $q(j)\ge 0$ necessitates
\be
\frac{\alpha}{\beta}\ge\frac{\sum_i r(j|i)}{\sum_i p(i)r(j|i)}\ge 1\;.
\ee

As before, we now start studying the consequences of the full requirement that
$0\le q(j)\le 1$, in the form of a generalized fundamental inequality:
\be
0\;\le\; \alpha\sum_{i=1}^n p(i)r(j|i) - \beta \sum_{i=1}^n r(j|i)\;\le\; 1\;.
\label{UrBoy}
\ee The two sets $\mathcal P$ and $\mathcal R$ are defined analogously to the
discussion just after Eq.~(\ref{TheTrueUr}), the first a set of priors for the
sky and the second a set of conditionals for the ground (given the outcomes $i$
in the sky).  $\mathcal P$ and $\mathcal R$ are taken to be consistent and
maximal.

Two assumptions, we shall borrow straight away from our previous development in Section \ref{Heimlich}.
\begin{resump}{\rm $\!\!$:}
{\rm Principle of Reciprocity:\ Posteriors from Maximal Ignorance Are Priors.}  For any $R\in\mathcal
R$, a posterior probability, consequent upon outcome $j$ of a ground
measurement, \be \mbox{\rm Prob}(i|j)=\frac{r(j|i)}{\sum_k r(j|k)}
\label{ChickenButt}
\ee may be taken as a valid prior $p(i)$ for the outcomes of the measurement in
the sky. Moreover, all valid priors $p(i)$ may arise in this way.
\end{resump}
\begin{resump}{\rm $\!\!$:}
{\rm Basis states span the simplex $\Delta_{d^2}$.}  The conditional
probabilities $r(j|i)$ derived from setting the ground measurement equal to the
sky measurement give rise to posterior distributions $\|e_k\drangle$, via
Eq.~(\ref{ChickenButt}), that span the whole probability simplex.
\end{resump}

At this stage, the argument goes just as it did in Section \ref{BengalTiger}.
In terms of the constants $\alpha$ and $\beta$, the
components $e_k(i)$ of the basis states satisfy the equations \be
e_k(i)=\frac{1}{\alpha}(\delta_{ki}+\beta)
\label{MorningBoy}
\ee
and
\be
\sum_i e_k(i)^2=\frac{1}{\alpha^2}\Big(1+2\beta+n\beta^2\Big)\;.
\label{CoffeeBelly}
\ee
Let us now consider a measurement with in-step unpredictability for a
measurement on the ground with $m$ outcomes ($m\ne n$)---that is, a measurement
on the ground such that if one has a flat distribution for the outcomes in the
sky, one will also have a flat distribution for the outcomes on the ground.
Let us again denote $r(j|i)$ by $r_{\rm \scriptscriptstyle ISU}(j|i)$ in this
special case. Following the manipulations we did before, we must have
\be
\sum_i r_{\rm \scriptscriptstyle ISU}(j|i) = \frac{n}{m}\;.
\ee
By the Principle of Reciprocity, this ISU measurement gives rise to a class of
priors which we denote by $\|p_k\drangle$, $k=1,\ldots,m$. Their components
are given by
\begin{equation}   \label{eq:ISUprior}
p_k(i)=\frac{m}{n} r_{\rm \scriptscriptstyle ISU}(k|i) \;;
\end{equation}
each vector $\|p_k\drangle$ represents a valid prior in the sky.

Let us now introduce a new notion that we did not make use of in the previous
development: We shall say that a measurement with in-step unpredictability {\rm
  achieves the ideal of certainty} if $\|p\drangle=\|p_k\drangle$ implies that
$q(j)=\delta_{jk}$, i.e., for such a measurement and a prior in the sky given
by $\|p_k\drangle$, the agent is certain that the outcome on the ground will be
$k$.

This leads to the following assumption:

\begin{resump}{\rm $\!\!$:} {\rm Availability of Certainty.}\footnote{In
    several axiomatic developments of quantum theory---see for instance
    \cite{Goyal08} and \cite{Hardy01}---the idea of repeated measurements
    giving rise to certainty (and the associated idea of ``distinguishable
    states'') is viewed as fundamental.  However, from the Quantum-Bayesian view
    where {\it all\/} measurements are generative of their outcomes---i.e.,
    outcomes never pre-exist the act of measurement---and certainty is always
    subjective certainty \cite{Caves07}, the consistency of adopting a state of
    certainty as one's state of belief, even in what is judged to be a repeated
    experiment, is not self-evident at all.  In fact, from this point of view,
    why one ever has certainty is the greater of the mysteries.}  For any
  system, there is a measurement with in-step unpredictability of some number
  $m_0\ge2$ of outcomes that (i) achieves the ideal of certainty and (ii) for
  which one of the priors $\|p_k\drangle$ defined in Eq.~(\ref{eq:ISUprior})
  has the form of a basis distribution (\ref{MorningBoy}).
\label{ass:certainty}
\end{resump}
For a measurement of this type, we have
\begin{equation}
\dlangle p_j\|
p_k\drangle=\frac{1}{\alpha}\left(\frac{m_0}{n}\delta_{jk}+\beta\right)\;,
\;\;\;\;j,k=1,\ldots,m_0\;,
\label{Delapidate}
\end{equation}
where $\dlangle \cdot\| \cdot\drangle$ denotes the inner
product. Using condition (ii) of the above resumption, it follows that
the squared norm $\dlangle p_k\|p_k\drangle$ of any of the vectors
$\|p_k\drangle$ is equal to the squared norm of the basis vectors given
by Eq.~(\ref{CoffeeBelly}). This, together with
Eq.~(\ref{RainyDay}) now implies that
\begin{equation}
\frac{m_0}{n}\alpha-\beta=1
\label{NimbleNooph}
\end{equation}
for any measurement satisfying Resumption~\ref{ass:certainty}.

Equation~(\ref{Delapidate}) expresses that any two of the vectors
$\|p_k\drangle$ differ by the same angle, $\theta$, defined by
\begin{equation}
\cos\theta = \frac{\dlangle p_1\|p_2\drangle}{\dlangle p_1\|p_1\drangle}\;.
\end{equation}
Using the relations~(\ref{RainyDay})
and~(\ref{NimbleNooph}) between our four variables, $\alpha$, $\beta$, $n$ and
$m_0$ established above, this angle can be seen to equal
\begin{equation}    \label{eq:gral}
\cos\theta = \frac{n-m_0}{(m_0-1)^2 + n-1} \;.
\end{equation}
We are now ready to state our last resumption.

\begin{resump}{\rm $\!\!$:} \label{ass:QBibbo} {\rm Many Systems, Universal Angle.}
The identity of a system is parameterized by its pair $(n,m_0)$. Nonetheless for all
systems, the angle $\theta$ between pairs of priors $\|p_k\drangle$ for any measurement satisfying
Resumption~\ref{ass:certainty} is a universal constant given by $\cos\theta=1/2$.
\end{resump}
The value $\cos\theta=1/2$ is less arbitrary than it may appear at first
sight. Taken by itself, the assumption that $\theta$ is universal implies
that, for any $m_0\ge2$, there is an integer $n$ such that the right-hand side of
Eq.~(\ref{eq:gral}) evaluates to the constant $\cos\theta$. It is not hard to
show that this is possible only if this constant is of the form
\begin{equation}
\cos\theta = \frac\qbar{\qbar+2} \;,
\end{equation}
where $\qbar$ is a non-negative integer. The universal angle postulated above
corresponds to the choice $\qbar=2$.

Every choice for $\qbar$ leads to a different relation between $n$ and $m_0$. For
$\qbar=0$, we find $n=m_0$, in which case the urgleichung turns out to be identical
to the classical law of total probability. For $\qbar=1$, we get the relationship
$n=\frac12 m_0(m_0+1)$ which, although this fact plays no role in our
argument, is characteristic of theories defined in real Hilbert space \cite{Wootters86}.
And for $\qbar=2$, we obtain
\begin{equation}   \label{eq:square}
n=m_0^2 \;.
\end{equation}

Equations (\ref{NimbleNooph}) and (\ref{eq:square}) hold for the special
measurement postulated in Resumption~\ref{ass:certainty}. If we eliminate
$m_0$ from these equations we find, with the help of Eq.~(\ref{RainyDay}), the
relationships
\begin{equation}
n = (\alpha-1)^2\;, \;\;\; \beta=\frac1{\sqrt n}\;.
\end{equation}
These equalities must hold for
{\it any\/} measurement on the ground. If we denote the integer $\alpha-1$ by
the letter $d$, we  recover the constants of the original
urgleichung of Eq.~(\ref{Reprise}).

Let us reiterate slightly the philosophy here.
The numerical relations between the constants $\alpha$, $\beta$, and $n$, and
in particular the fact that $n$ is a perfect square, follow from the
existence of a single special measurement defined in
Resumption~\ref{ass:certainty}, together with the postulate of a universal
angle in Resumption~\ref{ass:QBibbo}. These last two resumptions, as well as
the  first three, are given purely in terms of the personalist probabilities a
Bayesian agent may assign to the outcomes of certain experiments. Nowhere in
all this do we mention amplitudes, Hilbert space, or any other part of the
usual apparatus of quantum mechanics.

\section{Summary: From Quantum Interference to Quantum Bayesian Coherence}

In this paper, we hope to have given a new and useful way to think of quantum
interference: Particularly, we have shown how to view it as an empirical addition to
Dutch-book coherence, operative when one calculates probabilities for the
outcomes of a factualizable quantum experiment in terms of one explicitly
assumed counterfactual.  We did this and not once did we use the idea of a
probability {\it amplitude}.  Thus we believe we have brought the idea of
quantum interference formally much closer to its root in probabilistic
considerations. For this, we were aided by the mathematical machinery of SIC
measurements.

In so doing we showed that the Born Rule can be viewed as a relation
between probabilities, rather than a {\it setter\/} of probabilities from
something more firm or secure than probability itself, i.e., rather than
facilitating a probability assignment from {\it the\/} quantum state.  From
the Quantum-Bayesian point of view there is no such thing as {\it the\/}
quantum state, there being as many quantum states for a system as there are
agents interested in considering it.  This last point makes it particularly
clear why we needed a way of viewing the Born Rule as an extension of
Dutch-book coherence: One can easily invent situations where two agents will
update to divergent quantum states (even pure states, and even orthogonal pure
states, see Footnote \ref{NoConvergence})\ by looking at the same empirical
data \cite{Fuchs02,Fuchs04,Fuchs09}---a quantum state is always ultimately
dependent on the agent's priors.

But there is much more to do.  We gave an indication that the urgleichung and considerations to do with it already specifies a significant
fraction of the structure of quantum states---and for that reason one might
want to take it as one of the fundamental axioms of quantum mechanics.  We did
not, however, get all the way back to a set based on the manifold of pure
quantum states, Eq.~(\ref{MegaMorph}). A further open question concerns the
origin of the urgleichung. An intriguing idea would be to justify it
Dutch-book style in terms of bought and returned lottery tickets consequent
upon the nullification step in our standard scenario.  Then the positive
content of the Born Rule might be viewed as a kind of cost excised whenever one
factualizes a SIC.  But this is just speculation.

What is firm is that we have a new setting for quantifying the old idea that,
in quantum mechanics, unperformed measurements have no outcomes.

\section{Outlook}

\begin{flushright}
\baselineskip=13pt
\parbox{2.8in}{\baselineskip=13pt\small Of every would be describer of the
  universe one has a right to ask immediately two general questions.  The first
  is: ``What are the materials of your universe's composition?'' And the
  second: ``In what manner or manners do you represent them to be connected?''}
\medskip\\
\small --- William James, notebook entry, 1903 or 1904
\end{flushright}

This paper has focussed on adding a new girder to the developing structure of
Quantum Bayesianism (`QBism' hereafter).  As such, we have taken much of the
previously developed program as a background for the present efforts.  The core
arguments for why we choose a more `personalist Bayesianism' rather than a
so-called `objective Bayesianism' can be found in \cite{Fuchs02,Fuchs04,Struggles}.  The argument for why a subjective,
personalist account of certainty is crucial for breaking the impasse set by the
EPR criterion of reality are explained in \cite{Caves07,Struggles}.
Similarly for other questions on the program.

Still, fearing James' injunction, in this Section we want to discuss anew the
term `measurement,' which we have been using uncritically in the present paper.
Providing a deeper understanding of the proclamation `Unperformed measurements
have no outcomes!'\ is, we feel, the first step toward characterizing ``the
materials of our universe's composition.''

We take our cue from John Bell. Despite our liberal use of the term so far, we
think the word `measurement' should indeed be banished from fundamental
discussions of quantum theory~\cite{Bell90}.\footnote{For an argument in some
  sympathy with our own, see
  N.~D. Mermin's ``In Praise of Measurement'' \cite{Mermin06}.}  However, it is
not because the word is ``unprofessionally vague and ambiguous,'' as
Bell said of it~\cite{Bell87}.\footnote{To be sure, the are plenty of things
  vague and inconsistent in the writings of Bohr, Pauli, Heisenberg, von
  Weizs\"acker, Peierls, and Peres (representatives
  of the so-called `orthodoxy'), but this word we believe is not one of them.}  Rather, it is because, from the QBist
perspective, the word suggests a misleading notion of the very subject matter
of quantum mechanics.

To make the point dramatic, let us put quantum theory to the side for a moment,
and consider instead basic Bayesian probability theory.  There the subject
matter is an agent's expectations for various outcomes.  For instance, an agent
might write down a joint probability distribution $P(h_i,d_j)$ for various
mutually exclusive hypotheses $h_i$, $i=1,\ldots,n$, and data values $d_j$,
$j=1,\ldots,m$, appropriate to some phenomenon.  As discussed above,
a major role of the theory is that it provides a scheme
(Dutch-book coherence) for how these probabilities should be related to other
probabilities, $P(h_i)$ and $P(d_j)$ say, as well as to any other degrees of
belief the agent has for other phenomena.  The theory also prescribes that if
the agent is given a specific data value $d_j$, he should update his
expectations for everything else within his interest.  For instance, under the
right conditions \cite{Diaconis82,Skyrms87b,FuchsSchack2012}, he should reassess his
probabilities for the $h_i$ by conditionalizing: \be
P_{\mbox{new}}(h_i)=\frac{P(h_i,d_j)}{P(d_j)}\;.  \ee But what is this phrase
``given a specific data value''?  What does it really mean in detail?
Shouldn't one specify a mechanism or at least a chain of logical and/or
physical connectives for how the raw fact signified by $d_j$ comes into the
field of the agent's consciousness?  And who is this ``agent'' reassessing his
probabilities anyway?  Indeed, what is the precise definition of an agent? How
would one know one when one sees one?  Can a dog be an agent? Or must it be a
person? Maybe it should be a person with a PhD?\footnote{Tongue-in-cheek
  reference to Bell again \cite{Bell90}.}

Probability theory has no chance of answering these questions because they are
not questions within the subject matter of the theory.  Within probability
theory, the notions of ``agent'' and ``given a data value'' are primitive and
irreducible. Guiding agents' decisions based on
data is what the whole theory is constructed for.
As such, agents and data are the highest elements within
the structure of probability theory---they are not to be constructed from it,
but rather the former are there to receive the theory's guidance, and the
latter are there to designate the world external to the agent.

QBism says that, if all of this is true of Bayesian probability
theory in general, it is true of quantum theory as well.  As the foundations of
probability theory dismisses the questions of where data comes from and what
constitutes an agent---these questions never even come to its attention---so can the foundations of quantum theory dismiss them
too.  This point is one of the strongest reasons for making the QBist move in the first place.

\begin{figure} %\leavevmode
\begin{center}
\includegraphics[height=2.6in]{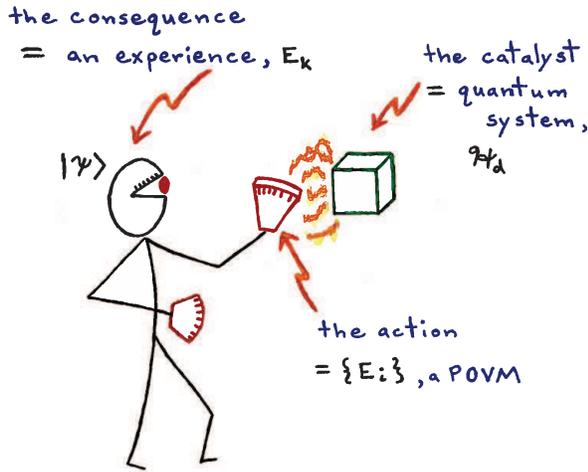}
%paulian2.eps}
\bigskip\caption{The Paulian Idea \cite{Fuchs01}---in the form of a figure
  inspired by John Archibald Wheeler \cite{WheelerFig}.  In contemplating a
  quantum measurement, one makes a conceptual
  split in the world: one part is treated as an agent, and the other as a kind
  of reagent or catalyst (one that brings about change in the agent).
  In older terms, the former is an observer and the latter a quantum system of
  some finite dimension $d$.  A quantum measurement consists first in the agent
  taking an {\it action\/} on the quantum system.  The action is formally
  captured by some POVM $\{E_i\}$. The action leads generally to an
  incompletely predictable {\it consequence}, a particular personal experience
  $E_i$ for the agent \cite{FuchsDelirium}.  The quantum state $|\psi\rangle$
  makes no appearance but in the agent's head; for it only captures his degrees
  of belief concerning the consequences of his actions, and---in contrast to
  the quantum system itself---has no existence in the external world.
  Measurement devices are depicted as prosthetic hands to make it clear that
  they should be considered an integral part of the agent.  (This contrasts
  with Bohr's view where the measurement device is always treated as a
  classically describable system external to the observer.)  The sparks between
  the measurement-device hand and the quantum system represent the idea that
  the consequence of each quantum measurement is a unique creation within the
  previously existing universe \cite{FuchsDelirium}.
  Wolfgang Pauli characterized this picture as a ``wider form of the
  reality concept'' than that of Einstein's, which he labeled ``the ideal of
  the detached observer'' \cite{Pauli94,Laurikainen88,Gieser05}. What is
  important for modern developments is that the particular character of the
  catalysts---i.e., James' ``materials of your universe's composition''---must
  leave its trace in the formal rules that allow us to conceptualize
  factualizable measurements in terms of a standard counterfactual one.}
\end{center}
\end{figure}

A likely reaction at this point will be along these lines: ``It
is one thing to say all this of probability theory, but quantum theory is a
wholly different story. Quantum mechanics is no simple branch of mathematics,
be it probability or statistics.  Nor can it plausibly be a theory about the
insignificant specks of life in our vast universe making gambles and
decisions. Quantum mechanics is one of our best theories of the world!  It is
one of the best maps we have drawn yet of what is actually out
there.''  But this is where QBism begs to differ. Quantum theory is {\it not\/} a
`theory of the world.'  Just like probability theory is not a theory of the
world, quantum theory is not as well: It is a theory for the use of agents
immersed in and interacting with a world of a particular character, ``the quantum
world.''

This last statement is crucial for understanding what we are trying
to say. Regarding the idea of a world external to
the agent, it must be as Martin Gardner says \cite{Gardner83}, \bq The
hypothesis that there is an external world, not dependent on human minds, made
of {\em something}, is so obviously useful and so strongly confirmed by
experience down through the ages that we can say without exaggerating that it
is better confirmed than any other empirical hypothesis.  So useful is the
posit that it is almost impossible for anyone except a madman or a professional
metaphysician to comprehend a reason for doubting it.  \eq Yet there is no
implication in these words that quantum theory, for all its success in
chemistry, physical astronomy, laser making, and so much else, must be read
as a theory of the world.  There is room for a significantly more interesting
form of dependence: Quantum theory is conditioned by
the character of the world, but yet is not a theory directly of it.  Confusion
on this very point, we believe, is what has caused most of the discomfort in
quantum foundations in the 85 years since the theory's coming to a relatively
stable form in 1927.

Returning to our discussion of Bell and the word ``measurement,'' we wish the
word banished because it subliminally whispers the philosophy of its
birth: That quantum mechanics should be conceived in a way that makes
no ultimate reference to agency, and that agents are constructed out of the
theory, rather than taken as the primitive entities the theory is meant to aid.
In a nutshell, the word deviously carries forward the impression that quantum
mechanics should be viewed as a theory directly of the world.

Fixing the word ``measurement'' is the prerequisite to a new ontology---in other words,
prerequisite to a statement about the (hypothesized) character of the world
that does not make direct reference to our actions and gambles within it.
Therefore, as a start, let us rebuild quantum mechanics in terms more conducive
to the Quantum Bayesian programme.

\subsection{The Paulian Idea and the Jamesian Pluriverse}

The best way to begin a more thoroughly QBist delineation of quantum mechanics
is to start with two quotes on personalist Bayesianism itself.  The first is
from Hampton, Moore, and Thomas \cite{Hampton73}, \bq Bruno de Finetti believes
there is no need to assume that the probability of some event has a uniquely
determinable value.  His philosophical view of probability is that it expresses
the feeling of an individual and cannot have meaning except in relation to him.
\eq and the second from D.~V. Lindley \cite{Lindley82}, \bq The Bayesian,
subjectivist, or coherent, paradigm is egocentric.  It is a tale of one person
contemplating the world and not wishing to be stupid (technically, incoherent).
He realizes that to do this his statements of uncertainty must be
probabilistic.  \eq These two quotes make it clear that personalist
Bayesianism is a ``single-user theory.''  Thus, QBism must inherit at least
this much egocentrism in its view of quantum states $\rho$.
The ``Paulian Idea'' \cite{Fuchs01}---which is also
essential to the QBist view---goes further still.  It says that the outcomes to
quantum measurements are single-user as well.  That is to say, when an agent
writes down her degrees of belief for the outcomes of a quantum measurement,
what she is writing down are her degrees of belief about her potential {\it
  personal\/} experiences arising in consequence of her actions upon the
external world \cite{FuchsDelirium,Mermin12,Fuchs10,Fuchs12}.

Before exploring this further, let us partially formalize in a quick outline
the structure of quantum mechanics from this point of view, at the moment
retaining the usual mathematical formulation of the theory, but starting the
process of changing the English descriptions of what the term ``quantum
measurement'' means.

\begin{enumerate}
\item
Primitive notions: a) the agent, b) things external to the agent, or, more
commonly, ``systems,'' c) the agent's actions on the systems, and d) the
consequences of those actions for her experience.

\item
The formal structure of quantum mechanics is a theory of how the agent ought to
organize her Bayesian probabilities for the consequences of all her potential
actions on the things around her.  Implicit in this is a theory of the
structure of actions.  The theory works as follows:

\item
When the agent posits a system, she posits a Hilbert space ${\mathcal H}_d$ as the arena for all her considerations.

\item
Actions upon the system are captured by positive-operator valued measures
$\{E_i\}$ on ${\mathcal H}_d$.  Potential consequences of the action are
labeled by the individual elements $E_i$ within the set.\footnote{There is a
  formal similarity between this and the development in Cox \cite{Cox61}, where
  ``questions'' are treated as sets, and ``answers'' are treated as elements
  within the sets.}  I.e.,
$$
\mbox{action}=\{E_i\} \quad \mbox{and} \quad \mbox{consequence}= E_k\;.
$$

\item
Quantum mechanics organizes the agent's beliefs by saying that she should
strive to find a single density operator $\rho$ such that her degrees of belief
will always satisfy \bea
\mbox{Prob}\Big(\mbox{consequence}\,\Big|\,\mbox{action}\Big)&=&
\mbox{Prob}\Big(E_k\,\Big|\,\{E_i\}\Big) \nonumber\\ &=&\tr\rho E_k\,,\nonumber
\eea no matter what action $\{E_i\}$ is under consideration.

\item
Unitary time evolution and more general quantum operations (completely positive
maps) do not represent objective underlying dynamics, but rather address the
agent's belief changes accompanying the flow of time, as well as belief changes
consequent upon any actions taken.

\item
When the agent posits {\it two\/} things external to herself, the arena for all
her considerations becomes ${\mathcal H}_{d_1}\otimes{\mathcal H}_{d_2}$.
Actions and consequences now become POVMs on ${\mathcal
  H}_{d_1}\otimes{\mathcal H}_{d_2}$.

\item
The agent can nonetheless isolate the notion of an action on a single one of
the things alone: These are POVMs of the from $\{E_i\otimes I\}$, and similarly
with the systems reversed $\{I \otimes E_i\}$.

\item
Resolving the consequence of an action on {\it one\/} of the things may cause
the agent to update her expectations for the consequences of any further
actions she might take on the {\it other\/} thing.  But for those latter
consequences to come about, she must elicit them through an actual action on
the second system.
\end{enumerate}

The present paper, of course, has predominantly focussed on Item 5 in this
list, rewriting the point in purely probabilistic terms.  With regard to the
discussion in the present Section, however, the main points to note are Items
4, 7, 8, and 9.  Regarding our usage of the word ``measurement,'' they say that
one should think of it simply as an {\it action\/} upon the system of interest.
Actions lead to consequences within the experience of the agent, and that is
what a quantum measurement is.  A quantum measurement finds nothing,
but very much {\it makes\/} something.

Thus, in a QBist painting of quantum mechanics, quantum measurements are
``generative'' in a very real sense.  But by that turn, the consequences of our
actions on physical systems must be egocentric as well.
Measurement outcomes come about for the agent himself. Quantum mechanics is a
single-user theory through and through---first in the usual Bayesian sense with
regard to personal beliefs, and second in that quantum measurement outcomes are
wholly personal experiences.

Of course, as a single-user theory, quantum mechanics is available to any agent
to guide and better prepare him for his own encounters with the world. And
although quantum mechanics has nothing to say about another agent's personal
experiences, agents can communicate and use the information gained from each
other to update their probability assignments. In the spirit of the Paulian
Idea, however, querying another agent means taking an action on him. Whenever ``I''
encounter a quantum system, and take an action upon it, it catalyzes a
consequence in my experience that my experience could not have foreseen.
Similarly, by a Copernican-style principle, I should assume the same for ``you'':
Whenever you encounter a quantum system, taking an action upon it, it catalyzes
a consequence in your experience.  By one category of thought, we are agents,
but by another category of thought we are physical systems.  And when we take
actions upon each other, the category distinctions are symmetrical.  Like with
the Rubin vase, the best the eye can do is flit back and forth between the two
formulations.

The previous paragraph should have made clear that viewing quantum mechanics
as a single user theory does not mean there is only one user. QBism does not lead
to solipsism.  Any charge of solipsism is further refuted by two points central
to the Paulian Idea. \cite{Fuchs02b}.  One is the conceptual split of the world
into two parts---one an agent and the other an external quantum system---that
gets the discussion of quantum measurement off the ground in the first place.
If such a split were not needed for making sense of the question of actions
(actions upon what?\ in what?\ with respect to what?), it would not have
been made. Imagining a quantum measurement without an autonomous quantum system
participating in the process would be as paradoxical as the Zen koan of the
sound of a single hand clapping.  The second point is that once the agent
chooses an action $\{E_i\}$, the particular consequence $E_k$ of it is beyond
his control.  That is to say, the particular outcome of a quantum measurement
is not a product of his desires, whims, or fancies---this is the very reason he
uses the calculus of probabilities in the first place: they quantify his
uncertainty \cite{Lindley06}, an uncertainty that, try as he might, he cannot
get around.  So, implicit in this whole picture---this whole Paulian Idea---is
an ``external world \ldots\ made of {\it something},'' just as Martin Gardner
calls for.  It is only that quantum theory is a rather small theory: Its
boundaries are set by being a handbook for agents immersed within that ``world
made of {\it something}.''

But a small theory can still have grand import, and quantum mechanics most
certainly does.  This is because it tells us how a user of the theory sees his
role in the world.  Even if quantum mechanics---viewed as an addition to
probability theory---is not a theory of the world itself, it is certainly
conditioned by the particular character of this world. Its empirical content
is exemplified by the simplest case of the urgleichung,
$$
q(j)=(d+1)\sum_{i=1}^{d^2} p(i) r(j|i) - 1\;,
$$
which takes this specific form rather than an infinity of other
possibilities.
Even though quantum theory is now understood as a theory of acts, decisions,
and consequences \cite{Savage54}, it tells us, in code, about the character of
our particular world.  Apparently, the world is made of a stuff that does not
have ``consequences'' waiting around to fulfill our ``actions''---it is a world
in which the consequences are generated on the fly.  One starts to get a sense
of a world picture that is part personal---{\it truly\/} personal---and part
the joint product of all that interacts.  This is a world of indeterminism, but
one with no place for ``objective chance'' in the sense of Lewis' Principal
Principle \cite{Lewis86a,Harper12}.  From within any part, the future is
undetermined.  If one of those parts is an agent, then it is an agent in a
situation of uncertainty.  And where there is uncertainty, agents should use
the calculus of Bayesian probability in order to make the best go at things.

\section{Acknowledgements}
We thank D. M. Appleby, G.~Bacciagaluppi, H.~C. von Baeyer, H.~Barnum,
R.~Blume-Kohout, H.~R. Brown, W.~Demopoulos, C.~Ferrie, S.~T. Flammia, L.~Hardy, P.~Hayden,
K.~Martin, N.~D. Mermin, R.~Morris, W.~C. Myrvold, D.~Parker, R.~W. Spekkens,
C.~Ududec, and J.~Uffink for discussions, and we especially thank {\AA}sa~Ericsson, Blake Stacey, and the anonymous referees for
many improvements to the evolving manuscript. \medskip

This research was supported in part by the U.~S. Office of Naval
Research (Grant No.\ N00014-09-1-0247). Research at PI is supported by the
Government of Canada through Industry Canada and by the Province of Ontario
through the Ministry of Research \& Innovation.

\end{document}